\documentclass[12pt]{article}
\usepackage{epsf}
\usepackage{amssymb}
\usepackage{latexsym}

\usepackage{graphicx,color}
\usepackage{wrapfig}
\usepackage{latexsym}
\usepackage{graphics}
\usepackage{amsmath} 
\usepackage{url}

\def\IN{\mathbb {N}}

\def\IC{\mathbb {C}}

\usepackage{color}
\usepackage{delarray,color}
\usepackage{fancybox} 
\usepackage{tabularx}

\catcode`\@=11
\@addtoreset{equation}{section}
\def\theequation{\thesection.\arabic{equation}}
\catcode`@=12
\relax

\makeatletter
\def\eqnarray{%
 \stepcounter{equation}%
 \let\@currentlabel=\theequation
 \global\@eqnswtrue
 \global\@eqcnt\z@
 \tabskip\@centering
 \let\\=\@eqncr
 $$\halign to \displaywidth\bgroup\@eqnsel\hskip\@centering
 $\displaystyle\tabskip\z@{##}$&\global\@eqcnt\@ne
 \hfil$\displaystyle{{}##{}}$\hfil
 &\global\@eqcnt\tw@$\displaystyle\tabskip\z@{##}$\hfil
 \tabskip\@centering&\llap{##}\tabskip\z@\cr}
\makeatother

\begin{document}
\baselineskip 17.5pt

\begin{titlepage}

\setcounter{page}{0}

\renewcommand{\thefootnote}{\fnsymbol{footnote}}
\vspace{-1cm}
\begin{flushright}
ITFA-2010-05
\end{flushright}

\vskip 1cm

\begin{center}
{\Large \bf 
The Volume Conjecture, \\[2mm]
Perturbative Knot Invariants, \\[2mm]
and \\[2mm]
Recursion Relations for Topological Strings
}

\vskip 1cm

{\normalsize
Robbert Dijkgraaf$^{1}$\footnote{r.h.dijkgraaf@uva.nl},
Hiroyuki Fuji$^{2}$\footnote{fuji@th.phys.nagoya-u.ac.jp}
and
Masahide Manabe$^{3}$\footnote{d07002p@math.nagoya-u.ac.jp}
}

\vskip 0.3cm

{ \it
$^1$ Institute for Theoretical Physics $\&$ KdV Institute
 for Mathematics\\
University of Amsterdam, Spui 21, 1012 WX Amsterdam, 
The Netherlands 
\\[2mm]
$^2$ Department of Physics, Nagoya University, Nagoya 464-8602, Japan
\\[2mm]
$^3$ Graduate School of Mathematics, Nagoya University, 
Nagoya 464-8602, Japan
}

\end{center}

\vspace{20mm}

\centerline{{\bf Abstract}} We study the relation between perturbative
knot invariants and the free energies defined by topological string
theory on the character variety of the knot. Such a correspondence
between $SL(2;\mathbb{C})$ Chern-Simons gauge theory and the
topological open string theory was proposed earlier on the basis of
the volume conjecture and AJ conjecture.  In this paper we discuss
this correspondence beyond the subleading order in the perturbative
expansion on both sides.  In the computation of the perturbative
invariants for the hyperbolic 3-manifold, we adopt the state integral
model for the hyperbolic knots, and the factorized AJ conjecture for
the torus knots.  On the other hand, we iteratively compute the free
energies on the character variety using the Eynard-Orantin topological
recursion relation.  We check the correspondence for the figure eight
knot complement and the once punctured torus bundle over
$\mathbb{S}^1$ with the holonomy $L^2R$ up to the fourth order.  For the
torus knots, we find trivial the recursion relations on both sides.

\end{titlepage}
\newpage

\section{Introduction}
Three-dimensional Chern-Simons gauge theory is one of the most widely
studied topological quantum field theories. It has found many
applications in physics and mathematics. In the celebrated paper by
Witten \cite{Witten_CS} a relation between the Chern-Simons gauge
theory and knot invariants was discovered, and it was shown that the
expectation value of the Wilson loop operator along the knot on
3-sphere $\mathbb{S}^3$ and the colored Jones polynomial are
equivalent.

Another remarkable aspect of the Chern-Simons gauge theory is its relation
with theories of three-dimensional quantum gravity \cite{Witten_grav}.
When gravity in three dimensions is reformulated in the first order
formalism, it also yields a Chern-Simons gauge theory.  In particular, in
this approach $SL(2;\mathbb{C})$ Chern-Simons gauge theory becomes
equivalent to the Euclidean signature quantum gravity with a negative
cosmological constant. In the classical limit this corresponds to the
study of the hyperbolic structures. Connecting these two aspects of
the Chern-Simons gauge theory, led to the volume conjecture as originally
proposed by Kashaev \cite{Kashaev}.

The volume conjecture concerns the asymptotic behavior of the colored
Jones polynomial \cite{Murakami^2}.  Let $J_n(K;q)$ be the $n$-colored
Jones polynomial for a hyperbolic knot $K$. The claim of the volume
conjecture is \cite{Kashaev,Murakami^2}:
\begin{eqnarray}
\frac{1}{\pi}\lim_{n\to \infty}\frac{|\log J_n(K;q=e^{2\pi i/n})|}{n}={\rm
 Vol}(\mathbb{S}^3\backslash K),
\end{eqnarray}
where ${\rm Vol}(\mathbb{S}^3\backslash K)$ is the hyperbolic volume
of the knot complement. The volume conjecture has been extended to the
complexified version \cite{Murakami^2}, and further generalized to
knot complements with the deformed hyperbolic structure \cite{Gukov}.
The volume conjecture beyond the leading order was discussed firstly
in \cite{GM}.  The subleading term in the asymptotic expansion of the
colored Jones polynomial coincides with the Reidemeister torsion
\cite{Porti}.

In previous work \cite{DijFu}, two of the authors proposed a
correspondence between $SL(2;\mathbb{C})$ Chern-Simons gauge
theory and the topological open string theory. This correspondence was
suggested by a similar set-up in the two problems. Let us briefly
review this argument. 

If $M$ is the three-manifold obtained by removing a tubular
neighbourhood of the knot $K$, then any quantum field theory on $M$
will produce a quantum state $Z(M)$ in the Hilbert space associated to
the boundary $\partial M$ of $M$. In the case of a knot complement,
the boundary has the topology $\partial M \cong
\mathbb{T}^2$. Semi-classically, the state $Z(M)$ is described by a Lagrangian
sub-manifold in the phase space associated to the boundary. For 
$SL(2;\mathbb{C})$ Chern-Simons gauge theory the classical phase space
can be identified with the space of gauge equivalence classes of the flat
$SL(2;\mathbb{C})$ connections on $\mathbb{T}^2$. The Lagrangian corresponding
to $M$ is the character variety $\cal C$ of the knot $K$, defined as
the set of connections on $\mathbb{T}^2$ that extend as flat connections over
$M$ \cite{CCGLS}. The character variety is an algebraic curve in (a
quotient of) $\mathbb{C}^* \times \mathbb{C}^*$ equipped with its
canonical symplectic structure. We can pick a local coordinate $u$ on
the curve $\cal C$, which can be identified with the free monodromy
around the meridian of the knot. The full CS partition function or
knot invariant now corresponds to the full quantum wave function
$Z(M;u)$.

This set-up of an algebraic curve $\cal{C} \subset \mathbb{C}^*
\times \mathbb{C}^*$ appears also in the topological string theory. In
this case we considered the toric Calabi-Yau 3-fold $X$ which can be regarded as a
fibration over the character variety of the knot. In this case the Riemann surface 
is usually called the spectral curve. If we add a topological
D-brane in this CY variety, the open string partition function will
also be a wave function $Z(X;u)$. So, in both the Chern-Simons gauge theory
and the topological string theory
we quantize the character variety. The conjecture that we study
further in this paper is that, with a suitable identification, these quantizations 
are equivalent, {\it i.e.}, we have
\begin{eqnarray}
Z(M;u) = Z(X;u).
\label{penta0}
\end{eqnarray}
This correspondence is summarized in more detail in the following
table. The various notations that we use here, will become clear in
the subsequent.

\begin{center}
\begin{tabular}{|c|c|}
\hline
{\bf 3D Chern-Simons} & {\bf  Topological Open String} \\ \hline
$u=2\pi i \left(\frac{n}{k}-1\right)$: Meridian holonomy  &
 $u$: Area of holomorphic disk \\
$q=e^{2\pi i/k}$ & $q=e^{g_s}$ \\
Vol+$i$CS &
Disk free energy ${\cal F}^{(0,1)}$ \\
Reidemeister torsion &
Annulus free energy ${\cal F}^{(0,2)}$ 
\\
AJ conjecture & Quantum Riemann Surface \\
${A}_K(\hat{m},\hat{\ell};q)J_n(K;q)=0$ & 
 $\hat{H}(e^{-u-g_s/2},e^{g_s\partial_u};q)Z_{\rm open}(u;q)=0$ \\
$\hat{m}\hat{\ell}=q^{1/2}\hat{\ell}\hat{m}$ & 
$e^{-u-g_s/2}e^{g_s\partial_u}=q^{1/2}e^{g_s\partial_u}e^{-u-g_s/2}$ \\
\hline
\end{tabular}
\end{center}

This correspondence is checked explicitly for two examples, the
figure eight knot complement and the once punctured torus bundle over
$\mathbb{S}^2$ with the holonomy $L^2R$, which is isomorphic to 
the SnapPea census manifold $m009$ \cite{HW}, 
which is the complement of a knot
in a three-manifold of different topology than the three-sphere
\cite{BR, magic, FKP}.  Up
to subleading order, one can find coincidence for these examples.

On the other hand, in \cite{Marino,BKMP} it is conjectured that the
free energies of the topological (A-type) open string theory on toric
Calabi-Yau 3-folds with toric branes are iteratively obtained by the
Eynard-Orantin topological recursion relation \cite{EO}. The recursion
relation is applicable for any complex plane curve, and in the
topological string theory, via the mirror symmetry, the open 
string moduli are described by the mirror curve which is a complex
plane curve in ${\IC}^* \times {\IC}^*$. These recursions can be derived
as the Schwinger-Dyson equations in two dimensional Kodaira-Spencer
theory, the theory of a chiral boson on the mirror curve
\cite{DVKS}. The techniques of the computation are developed in
\cite{Bouchard:2008gu,Brini,Manabe}. In this paper, we discuss our
correspondence to the higher orders beyond the subleading order in the
topological expansion of the free energies of the topological open
string theory. To this end we define the BKMP's free energies according to
remodeling the B-model \cite{Marino, BKMP}, and compute them for the
character variety of the hyperbolic manifold up to the fifth order in
the recursions, in the case of the above two examples.

On the Chern-Simons gauge theory side, the higher order perturbative
invariants are computed in \cite{DGLZ,DG}.  For the figure eight knot
complement we can compute the perturbative invariant from the colored
Jones polynomial. But for the once punctured torus bundle the complete
form of the colored Jones polynomial is not known.  To analyze such
manifolds, we adopt the state integral model which is constructed in
\cite{Hikami1,Hikami2,Hikami_gen,DGLZ}.  The partition function of the
state integral model for a simplicially decomposed hyperbolic
3-manifold gives topological invariants like the Ponzano-Regge
\cite{PR} and the Turaev-Viro models \cite{TV}.  In this paper, we
compute the perturbative invariants from the state integral model, and
compare them with the free energy on the character variety.

As a result of these computations, we find some discrepancies between
the perturbative invariants and the BKMP's free energy of the
topological string in the higher order.  These discrepancies may come from
the choice of the integration path in the computation of the free energy 
in the B-model or ${\cal O}(g_s)$ modification of the Calabi-Yau geometry 
\cite{KPW,DGH,AFKMY}.  To remedy this point, we consider some 
regularization for the constant $G$ which appears in the Bergman kernel, 
because the constant $G$ changes under the monodromy transformation of 
the genus one character variety.  
Although the regularization would be ad-hoc, we find the regularization
rules for $G^n$ terms in the recursions up to the fourth order.
After the regularization, we recover the perturbative invariants
of the state integral model for the above two examples non-trivially. 

For the case of torus knots, the colored Jones polynomial is
well-studied. We can extract the perturbative invariants adopting the
$q$-difference equation which is called the AJ conjecture
\cite{Garou1,Garou2, Garou-Le,Geronimo,HikamiAJ}.  Compared to the
state integral model, there exist two branches which correspond to
abelian and non-abelian representations of the $PSL(2;\mathbb{C})$
holonomy along the meridian for the colored Jones polynomial.  
In this paper we will compare the perturbative invariants for 
the non-abelian branch and the BKMP's free
energy on the character variety.  In this class of knots we find that
the perturbative invariants and BKMP's free energies are trivial as
the $u$-dependent functions, and we can check our correspondence 
to all orders in the perturbative expansion.

The organization of this paper is as follows.  In section 2, after a
short summary of the state integral model, we show the explicit
computation for the figure eight knot complement and the once
punctured torus bundle over $\mathbb{S}^1$ with the holonomy $L^2R$.
The computation of the figure eight knot complement was already given
in \cite{DGLZ}, and the second example is novel.  In section 3, we
turn to the computation of the BKMP's free energies on the character
variety. In this section, we firstly derive the general solution of
the topological recursion relations for the two branched plane curve with genus
one up to the fourth order.  And then we apply the formula to the
character varieties for the figure eight knot complement and the once
punctured torus bundle over $\mathbb{S}^1$ with the holonomy
$L^2R$. The correspondence for the torus knots is discussed in section
4.  In appendix A, we discuss the AJ conjecture.  We explicitly see the
factorization of the $q$-difference equation for the figure eight knot,
and summarize the computation of the perturbative invariants for some
torus knots in the abelian branch.  In appendix B, we summarize the
details of the derivations of the general formula for the fourth order
terms $W^{(0,4)}$ and $W^{(1,2)}$.  In appendix C, we show the
computation on the annulus free energy ${\cal F}_1(p)$. In appendix D,
the computational result of the fifth order free energy ${\cal
  F}_4(p)$ is summarized.

\section{Asymptotic expansion for the state integral model}
The volume conjecture describes the asymptotic behavior of the colored
Jones polynomial. To evaluate the saddle point value perturbatively,
the information of the cyclotomic expansion of the colored Jones
polynomial is necessary because we use a $q$-difference equation in
the analysis.  In general it is not an easy task to obtain such an
expansion.  In particular for the once punctured torus bundle over
$\mathbb{S}^1$, the original colored Jones polynomial can not be
found, although the simplicial decomposition is realized explicitly.

The partition function of the state integral model
\cite{Hikami1,Hikami2,Hikami_gen} on the hyperbolic $3$-manifold gives
the topological invariant.  The asymptotic expansion of the partition
function is computable only from the information of the simplicial
decomposition via ideal tetrahedra.
For the figure eight knot, the asymptotic expansions of the partition
function of the state integral model and the colored Jones polynomial
are computed thoroughly in \cite{DGLZ}, and both of them give the same
expansions.  In this section, we evaluate the asymptotic expansion of
the state integral model based on the method which is developed in
\cite{DGLZ}.

\subsection{State integral model}
Here we briefly summarize about the state integral model.
For a hyperbolic knot complement, the simplicial decomposition 
with the ideal tetrahedra can be performed
\cite{Thurston}.
There are two kinds of ideal tetrahedra with an orientation
$\varepsilon=\pm 1$.
For each face of the tetrahedron, a vector space $V$
or its dual $V^{*}$ is assigned corresponding to the orientation. 
The vector space $V$ is the Hilbert space of the Heisenberg algebra
with continuum eigenvalue for the momentum operator:
\begin{eqnarray}
&&[\hat{q},\hat{p}]=2\hbar, \\
&&\hat{p}|p\rangle=p|p\rangle,\quad  |p\rangle \in V.
\end{eqnarray}
Thus a tensor product of the 
vector spaces $V\otimes V\otimes V^*\otimes V^*$ is assigned
for an ideal tetrahedron.

As a weight of the state integral model, one can choose 
the matrix element of the 
operator ${\bf S}\in {\rm Hom}_{\mathbb{C}}(V\otimes V,V\otimes V)$. 
To acquire topological invariance
for the partition function, the operator should satisfy the pentagon
relation on $V_1\otimes V_2\otimes V_3$:
\begin{eqnarray}
{\bf S}_{2,3}{\bf S}_{1,2}={\bf S}_{1,2}{\bf S}_{1,3}{\bf S}_{2,3},
\label{penta}
\end{eqnarray}
where ${\bf S}_{i,j}$ acts on $V_i\otimes V_j$.
In state integral model \cite{Hikami1,Hikami_gen}, 
the operator ${\bf S}$ which satisfy the above pentagon
relation on the infinite dimensional momentum space $|p\rangle$ 
is defined on $V_1\otimes V_2$ as:
\begin{eqnarray}
&&{\bf S}_{1,2}=e^{\hat{q}_1\hat{p}_1/2\hbar}\Phi_{\hbar}
(\hat{p}_1+\hat{q}_2-\hat{p}_2),
\end{eqnarray}
where $\Phi_{\hbar}(p)$ is the quantum dilogarithm function. 
For $i\hbar\in \mathbb{R}$ and $|{\rm Im}\; p|<\pi$, the function 
$\Phi_{\hbar}(p)$ is given by the Faddeev's integral formula \cite{Faddeev}:
\begin{eqnarray}
\Phi_{\hbar}(p)=\exp\left[\frac{1}{4}\int_{\mathbb{R}+i0}
\frac{dx}{x}\frac{e^{-ipx}}{\sinh (i\hbar x)\sinh (\pi x)}\right].
\end{eqnarray}
Because this Faddeev integral fulfills the pentagon relation:
\begin{eqnarray}
\Phi_{\hbar}(\hat{p})\Phi_{\hbar}(\hat{q})
=\Phi_{\hbar}(\hat{q})\Phi_{\hbar}(\hat{p}+\hat{q})\Phi_{\hbar}(\hat{p}),
\end{eqnarray}
the pentagon relation (\ref{penta})
for ${\bf S}$ operators is implied.

\begin{figure}[t]
 \begin{center}
  \includegraphics[width=10cm,clip]{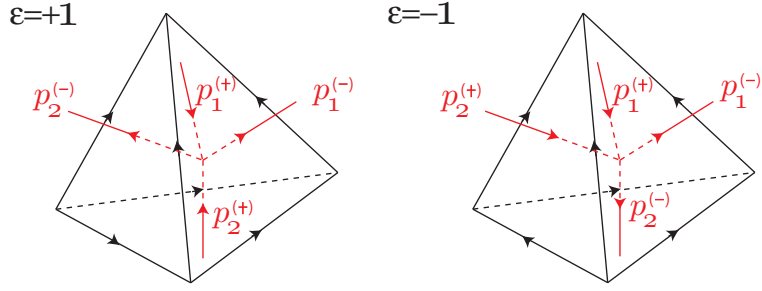}
 \end{center}
 \caption{Ideal tetrahedra of different orientation}
 \label{fig:tetra}
\end{figure}

The ${\bf S}$-matrix elements for the ideal tetrahedra of $\varepsilon=\pm 1$ as in Fig.\ref{fig:tetra} are obtained as follows:
\begin{eqnarray}
\langle p_1^{(-)},p_2^{(-)}|{\bf S}|p_1^{(+)},p_2^{(+)}\rangle
&=&\frac{\delta\big(p_1^{(-)}+p_2^{(-)}-p_1^{(+)}\big)}{\sqrt{4\pi
\hbar/i}}\Phi_{\hbar}(p_2^{(+)}-p_2^{(-)}+i\pi+\hbar)
\nonumber \\
&&\times
e^{\frac{1}{2\hbar}\left[p_1^{(-)}(p_2^{(+)}-p_2^{(-)})+\frac{i\pi\hbar}{2}-\frac{\pi^2-\hbar^2}{6}\right]}, 
\\
\langle p_1^{(-)},p_2^{(-)}|{\bf S}^{-1}|p_1^{(+)},p_2^{(+)}\rangle
&=&\frac{\delta\big(p_1^{(-)}-p_1^{(+)}-p_2^{(+)}\big)}{\sqrt{4\pi \hbar/i}}
\frac{1}{\Phi_{\hbar}(p_2^{(-)}-p_2^{(+)}-i\pi-\hbar)}
\nonumber \\
&&\times
e^{\frac{1}{2\hbar}\left[-p_1^{(+)}(p_2^{(-)}-p_2^{(+)})-\frac{i\pi\hbar}{2}+\frac{\pi^2-\hbar^2}{6}\right]}.
\end{eqnarray}

The partition function for the state integral model
on a 3-manifold $M$ with the complete
hyperbolic structure is defined by
\begin{eqnarray}
Z_{\hbar}(M;\hbar,0)=\sqrt{2}\int d{\bf p}\delta_C({\bf
 p})\delta_G({\bf p})\prod_{i=1}^N\langle
 p_{2i-1}^{(-)},p_{2i}^{(-)}|{\bf S}^{\varepsilon_i}|p_{2i-1}^{(+)},p_{2i}^{(+)}
\rangle, 
\end{eqnarray}
where ${\bf p}$ is a set of
$(p_{2i-1}^{(\varepsilon_i)},p_{2i}^{(\varepsilon_i)})$.  The delta
functions $\delta_G({\bf
  p}):=\prod\delta\big(p_j^{(-)}-p_k^{(+)}\big)$ and $\delta_C({\bf
  p})$ imply the gluing condition for the faces and the complete
gluing condition for the triangles on the boundary $\partial M\simeq
\mathbb{T}^2$ respectively. The asymptotic behavior of this invariant
is studied for various knot complements
\cite{Hikami1,Hikami_gen}.


The partition function for the 3-manifold with the 
deformed complete structure is also considered \cite{DGLZ}. 
The deformation of the completeness condition changes one of the delta
functions in $\delta_{C}({\bf p})$ with $\delta(\sum p_i=2u)$. 
In this delta function, the sum is taken for the momenta which appear 
in the shape parameters $z_i=e^{p_{2i}^{(+)}-p_{2i-1}^{(-)}}$ 
of the ideal tetrahedra along the meridian of the boundary torus 
\cite{Hikami2}. 
As a result of this deformation, one finds a holonomy representation along the meridian $\mu$:
\begin{eqnarray}
\rho(\mu)=\left(
\begin{array}{cc}
m & *~ \\
0 & m^{-1}
\end{array}
\right), \quad m=e^{u}.
\end{eqnarray}
Under this deformation, the partition function yields
\begin{eqnarray}
Z_{\hbar}(M;\hbar,u)&=&
\sqrt{2}e^{-u}\int d{\bf p}
\delta_C({\bf p};u)\delta_G({\bf p})\prod_{i=1}^N\langle
 p_{2i-1}^{(-)},p_{2i}^{(-)}|{\bf S}^{\varepsilon_i}|p_{2i-1}^{(+)},p_{2i}^{(+)}
\rangle
\nonumber \\
&=&
\sqrt{2}\int\prod_{j=1}^{N-1}\frac{dp_j}{\sqrt{4\pi\hbar}}
\prod_{i=1}^N\Phi_{\hbar}\left(
g_i({\bf p},2u)+\varepsilon_i(i\pi+\hbar)
\right)^{\varepsilon_i}e^{\frac{1}{2\hbar}f({\bf p},2u,\hbar)-u},
\label{state_integral} 
\end{eqnarray}
where $g_i({\bf p},2u)$ is a linear function of ${\bf p}$, and 
$f({\bf p},2u,\hbar)$ is a quadratic polynomial.

Although the contour of the integrations of the partition function is
not defined explicitly, the WKB expansion around the saddle point is
computable \cite{DGLZ}. 
In the limit $\hbar\to 0$, the partition function (\ref{state_integral})
is approximated by
\begin{eqnarray}
&&Z_{\hbar}(M;\hbar,u)\sim \int d{\bf p}\; e^{\frac{1}{\hbar}V({\bf
 p},u)},\quad \hbar\to 0,
\\
&&V({\bf p},u)=\frac{1}{2}\sum_{i=1}^N \varepsilon_i{\rm Li}_2
\left(-\exp(g_i({\bf p},2u)+i\pi \varepsilon_i)
\right)+\frac{1}{2}f({\bf p},2u,\hbar=0).
\end{eqnarray}
The saddle point conditions
\begin{eqnarray}
\frac{\partial V({\bf p},u)}{\partial p_j}=0
\label{saddle}
\end{eqnarray}
for all $j$ specify the saddle point ${\bf p}$.
There may exist several saddle points for a hyperbolic 3-manifold
$M$, and the 
value of the partition function may differ for each branch.
In the following, we discuss only the geometric branch.
For the general
saddle point values, the face angles of the ideal tetrahedra 
which is determined by a shape parameters may become non-geometric
\cite{Thurston, Boyd}. 
In the geometric branch, all of the ideal tetrahedra in the simplicial
decomposition are geometric and completely glued.\footnote{There
also exists the conjugate branch which satisfies $v^{(\rm
geom)}=-v^{(\rm conj)}$. In particular for the fully ampherical knot
 $S_{n+1}^{(\rm geom)}=(-1)^{n+1}S_{n+1}^{(\rm conj)}$ is also satisfied.} The saddle point value
of the potential $V({\bf p},u)$ will 
satisfy $V({\bf p}_0,u=0)=i{\rm Vol}(M)-{\rm CS}(M)$. 
This property is the same as the volume conjecture \cite{Kashaev,
Murakami^2}, and it is expected that the perturbative invariants for the
state integral model will coincide with those of colored Jones polynomial.

In \cite{Hikami_gen} it is proposed that
the potential function $V({\bf p}_0,u)$ is identified with the Neumann-Zagier
potential \cite{NZ}.
The Neumann-Zagier potential satisfies a relation:
\begin{eqnarray}
\frac{\partial}{\partial u}V({\bf p}_0,u)=v,
\label{NZ}
\end{eqnarray}
where $l:=e^{v}$ is an eigenvalue of the holonomy representation $\rho(\nu)$ along the longitude $\nu$ of the boundary $\partial M$.
This condition imposes a non-trivial constraint on $(l,m)\in
\mathbb{C}^*\times\mathbb{C}^*$. 
By the saddle point equation (\ref{saddle}) one can eliminate ${\bf p}$
and find an algebraic equation\footnote{In this algebraic equation, a
factor $l-1$ is not included. This indicates the absence of the abelian
branch in the state integral model \cite{DGLZ}.}:
\begin{eqnarray}
A_K(l,m)=0.
\end{eqnarray}
The polynomial $A_K(l,m)$ coincides with the A-polynomial which is
reciprocal $A_K(l,m)=l^am^bA_K(l^{-1},m^{-1})$. 
Then the saddle
point value of the potential $V({\bf p},u)$ will satisfy
 the generalized volume conjecture
 $V({\bf p}_0,u)=i{\rm Vol}(M,u)-{\rm CS}(M,u)$ \cite{Gukov}.

The higher order terms in the expansion around the saddle point is evaluated by
the following expansion of the quantum dilogarithm function:
\begin{eqnarray}
\Phi_{\hbar}(p_0+p)&=&\exp\bigg[
\sum_{n=0}^{\infty}B_n\Big(\frac{p}{2\hbar}+\frac{1}{2}\Big)
{\rm Li}_{2-n}(-e^{p_0})\frac{(2\hbar)^{n-1}}{n!}
\bigg],
\label{dilog_exp}
\end{eqnarray}
where the Bernoulli polynomial $B_n(x)=\sum_{k=0}^n {}_nC_kB_kx^{n-k}$
satisfies $B_n^{\prime}(x)=nB_{n-1}(x)$.
Plugging this expansion into (\ref{state_integral}), one can expand the 
partition function $Z_{\hbar}(M;\hbar,u)$ around the saddle point ${\bf p}_0$ as:
\begin{eqnarray}
Z_{\hbar}(M;\hbar,u)=\exp\bigg[\frac{1}{\hbar}S_0(u)+
\sum_{n=0}^{\infty}S_{n+1}(u)\hbar^n
\bigg], \quad 
 S_0(u)=V({\bf p}_0,u).
\end{eqnarray}
In the following, we will compute the higher order terms in the saddle
point approximation for the figure eight knot complement \cite{DGLZ} 
and the once puncture
torus bundle over $\mathbb{S}^1$ with the holonomy $L^2R$.

\subsection{Perturbative expansion for figure eight knot complement}

As the first example we summarize the computation for the figure
eight knot complement \cite{Hikami_gen, DGLZ}.\footnote{This computation
is already shown in \cite{DGLZ}.}
The figure eight knot complement can be decomposed into 
two ideal tetrahedra with the different orientations \cite{Thurston}.
The partition function for the state integral model yields
\begin{eqnarray}
Z_{\hbar}(\mathbb{S}^3\backslash{\bf 4_1};\hbar,u)=\sqrt{2}e^{-u}
\int d{\bf p}\delta_C({\bf p},u)\langle p_1,p_2|S|p_3,p_4\rangle
\langle p_4,p_3|S^{-1}|p_2,p_1\rangle.
\label{fig8_int}
\end{eqnarray}
The shape parameters $z_1=e^{p_4-p_2}$ and $z_2=e^{p_1-p_3}$
satisfies the meridian condition $z_1 z_2^{-1}=e^{-2u}$, and 
the delta function $\delta_C({\bf p};u)$ has the support on
\begin{eqnarray}
p_4-p_2-(p_1-p_3)=-2u.
\end{eqnarray}
Evaluating some of the integrals in (\ref{fig8_int})
one obtains
\begin{eqnarray}
Z_{\hbar}(\mathbb{S}^3\backslash{\bf 4_1};\hbar,u)
=\frac{1}{\sqrt{2\pi\hbar}}
\int_{C} dp\frac{\Phi_{\hbar}(p+i\pi+\hbar)}{\Phi_{\hbar}(-p-2u-i\pi-\hbar)}e^{-\frac{2}{\hbar}u(u+p)-u}.
\end{eqnarray}

Since the figure eight knot is fully amphichiral, the term $i\pi+\hbar$
in the argument of the quantum dilogarithm can be removed by shifting 
$p\to p-u-i\pi-\hbar$.
This shift of the variable changes the integration path ${C}$.
Although such shift gives rise to the corrections 
of order $e^{-{\rm const}/\hbar}$ to $Z_{\hbar}$, 
the higher order terms $S_{n}$ ($n=1,\ldots$) are not affected.

Under the shift of the variable $p\to p-u-i\pi-\hbar$, the partition
function of the state integral model simplifies
\begin{eqnarray}
Z(\mathbb{S}^3\backslash {\bf 4_1};\hbar,u)
&=&\frac{1}{\sqrt{2\pi\hbar}}e^{\frac{2\pi i u}{\hbar}+u}
\int_{C}dp\frac{\Phi_{\hbar}(p-u)}{\Phi_{\hbar}(-p-u)}\nonumber\\
&=&\frac{1}{\sqrt{2\pi\hbar}}e^{\frac{2\pi i u}{\hbar}+u}
\int_{C}dp\; e^{\Upsilon(\hbar,p,u)},
\end{eqnarray}
where one can compute the expansion of the function $\Upsilon(\hbar,p,u)$ 
adopting (\ref{dilog_exp}) as:
\begin{eqnarray}
&&\Upsilon(\hbar,p_0+p,u)=\sum_{j=0}^{\infty}\sum_{k=-1}^{\infty}
\Upsilon_{j,k}(p,u)p^j\hbar^k, \\
&& \Upsilon_{j,k}=\frac{B_{k+1}(1/2)2^k}{(k+1)!j!}\left[
{\rm Li}_{1-j-k}(-e^{p-u})-(-1)^j{\rm Li}_{1-j-k}(-e^{-p-u})
\right].
\end{eqnarray}
In the following, we will evaluate $Z(\mathbb{S}^3\backslash
{\bf 4_1};\hbar,u)$ on the geometric branch. In this branch, the saddle point
value of $p_0$ is
\begin{eqnarray}
&&p_0=p^{({\rm geom})}(u)=\log \bigg[\frac{1-m^2-m^4-m^2\Delta(m)}{2m^3}\bigg],
\\
&&  \Delta(m)=\sqrt{m^{-4}-2m^{-2}-1-2m^2+m^4}.
\end{eqnarray}

The perturbative invariants $S_{n}(u)$ are computed systematically by 
evaluating the Gaussian integrals,
\begin{equation}
Z_{\hbar}(\mathbb{S}^3\setminus {\bf 4_1};\hbar,u)
=\frac{e^{u+\frac{1}{\hbar}V(u)}}{\sqrt{2\hbar}}
\int_{C}dp\; e^{-\frac{b(u)}{2\hbar}p^2}\exp\bigg[
\frac{1}{\hbar}\sum_{j=3}^{\infty}\Upsilon_{j,-1}p^j
+\sum_{j=0}^{\infty}\sum_{k=1}^{\infty}\Upsilon_{j,k}\hbar^kp^j
\bigg].
\end{equation}
In the geometric branch, the first two terms yield
\begin{eqnarray}
&&S_{0}=V(u)
=\frac{1}{2}\Big[
{\rm Li}_2(-e^{p^{({\rm geom})}-u})-{\rm Li}_2(-e^{-p^{({\rm
geom})}-u})-4p^{({\rm geom})}u+4\pi i u
\Big],
\\
&& S_1=-\frac{1}{2}\log\frac{b(u)}{m^2}
=-\frac{1}{2}\log[i\Delta(m)/2],
\label{S1Fig}
\end{eqnarray}
The result $S_1(u)$ is consistent with the Reidemeister torsion of
$\mathbb{S}^3\backslash {\bf 4_1}$ \cite{Porti,GM}. 
The A-polynomial is computed from the equations
(\ref{saddle}) and (\ref{NZ})
with $S_0(u)=V(u)$,
\begin{eqnarray}
A_{\bf 4_1}(l,m)=-m^4+l(1-m^2-2m^4-m^6+m^8)-l^2m^4.
\label{AplyFig}
\end{eqnarray}

The higher order terms are computed in the same way:
\begin{eqnarray}
&&S_2=\frac{15}{2b^3}\Upsilon_{3,-1}^2+\frac{3}{b^2}\Upsilon_{4,-1}+\Upsilon_{0,1}, \\
&&S_3=
\frac{3465}{8b^6}\Upsilon_{3,-1}^4
+\frac{945}{2b^5}\Upsilon_{3,-1}^2\Upsilon_{4,-1}
+\frac{105}{2b^4}\left(
2\Upsilon_{3,-1}\Upsilon_{5,-1}+\Upsilon_{4,-1}^2\right)
\nonumber \\
&&
\hspace*{1cm}
+\frac{15}{2b^3}\left(
2\Upsilon_{6,-1}+\Upsilon_{0,1}\Upsilon_{3,-1}^2\right)
+\frac{3}{b^2}\left(\Upsilon_{0,1}\Upsilon_{4,-1}
+\Upsilon_{1,1}\Upsilon_{3,-1}\right)
\nonumber \\
&&\hspace*{1cm}
+\frac{1}{b}\Upsilon_{2,1}
+\Upsilon_{0,2}+\frac{1}{2}\Upsilon_{0,1}^2-\frac{S_2^2}{2},\\
&&S_4=\frac{1}{24b^7}\left(
810810\Upsilon_{3,-1}^2\Upsilon_{4,-1}^2 
+540540\Upsilon_{3,-1}^3\Upsilon_{5,-1}
\right)
\nonumber \\
&&\hspace*{1cm}
+\frac{3465}{8b^6}\left(
\Upsilon_{0,1}\Upsilon_{3,-1}^4+4\Upsilon_{4,-1}^3
+24\Upsilon_{3,-1}\Upsilon_{4,-1}\Upsilon_{5,-1}
+12\Upsilon_{3,-1}^2\Upsilon_{6,-1}\right)
\nonumber \\
&&\hspace*{1cm}
+\frac{315}{2b^5}\left(
\Upsilon_{1,1}\Upsilon_{3,-1}^3
+3\Upsilon_{0,1}\Upsilon_{3,-1}^2\Upsilon_{4,-1}
+3\Upsilon_{5,-1}^2+6\Upsilon_{4,-1}\Upsilon_{6,-1}
+6\Upsilon_{3,-1}\Upsilon_{7,-1}
\right)
\nonumber \\
&&\hspace*{1cm}
+\frac{105}{2b^4}\left(
\Upsilon_{2,1}\Upsilon_{3,-1}^2+2\Upsilon_{1,1}\Upsilon_{3,-1}\Upsilon_{4,-1}
+\Upsilon_{0,1}\Upsilon_{4,-1}^2+2\Upsilon_{0,1}\Upsilon_{3,-1}\Upsilon_{5,-1}
+2\Upsilon_{8,-1}
\right)
\nonumber \\
&&\hspace*{1cm}
+\frac{15}{4b^3}\left(
\Upsilon_{0,1}^2\Upsilon_{3,-1}^2+2\Upsilon_{0,2}\Upsilon_{3,-1}^2
+4\Upsilon_{3,-1}\Upsilon_{3,1}+4\Upsilon_{2,1}\Upsilon_{4,1}
+4\Upsilon_{1,1}\Upsilon_{5,-1}+4\Upsilon_{0,1}\Upsilon_{6,-1}
\right)
\nonumber \\
&&\hspace*{1cm}
+\frac{3}{2b^2}\left(
2\Upsilon_{0,1}\Upsilon_{1,1}\Upsilon_{3,-1}+2\Upsilon_{1,2}\Upsilon_{3,-1}
+\Upsilon_{0,1}^2\Upsilon_{4,-1}+2\Upsilon_{0,2}\Upsilon_{4,-1}
+2\Upsilon_{4,1}
\right)
\nonumber \\
&&\hspace*{1cm}
+\frac{1}{b}\left(\Upsilon_{1,1}^2+2\Upsilon_{0,1}\Upsilon_{2,1}
+2\Upsilon_{2,2}\right)
+\frac{1}{6}\left(\Upsilon_{0,1}^3+6\Upsilon_{0,1}\Upsilon_{0,2}
+6\Upsilon_{0,3}\right)-\frac{S_2^3}{6}-S_2S_3.
\end{eqnarray}

Applying this expansion, one finds the perturbative invariants in the
geometric branch yields
\begin{eqnarray}
\label{S2Fig}
S_2(u)
&=&\frac{-1}{12\Delta^3m^6}(1-m^2-2m^4+15m^6-2m^8-m^{10}+m^{12}),\\
\label{S3Fig}
S_3(u)
&=&\frac{2}{\Delta^6m^6}(1-m^2-2m^4+5m^{6}-2m^8-m^{10}+m^{12}),\\
\label{S4Fig}
S_4(u)
&=&\frac{1}{90\Delta^9m^{16}}(1-4m^2-128m^4+36m^6+1074m^8
-5630m^{10}+5782m^{12}
\nonumber \\ &&
+7484m^{14}
-18311m^{16}
+7484m^{18}+5782m^{20}-5630m^{22}+1074m^{24}
\nonumber \\ &&
+36m^{26}-128m^{28}-4m^{30}+m^{32}).
\end{eqnarray}
In \cite{DGLZ} the perturbative invariants are computed up to the eighth order.

\subsection{Perturbative expansion for once punctured torus bundle over
  $\mathbb{S}^1$ with holonomy $L^2 R$}

The next example is the once punctured torus bundle over $\mathbb{S}^1$. 
This class of manifolds is studied in the Jorgensen's theory on the space 
of quasifuchsian (once) punctured torus groups from the view point of 
their Ford fundamental domains \cite{Jorgensen,ASWY}.
In particular, the complete hyperbolic structure of this class of manifolds
is studied well, and the ideal triangulation is found explicitly
\cite{FH}.

Let $T_{\varphi}$ be a once punctured torus bundle over $\mathbb{S}^1$
\cite{Jorgensen}:
\begin{eqnarray}
&&T_{\varphi}=F\times I/\sim, \\
&& F=\mathbb{T}^2\backslash \{0\}, \quad I=[0,1],\quad (x,0)\sim (\varphi(x),1),
\nonumber 
\end{eqnarray}
where $T_{\varphi}$ admits a hyperbolic structure, if the monodromy
matrix $\varphi$ has two distinct eigenvalues \cite{Otal,GF}. 
Such a monodromy matrix
is specified by a sequence of $2p$ positive integers
$(a_1,b_1,a_2,b_2,\ldots a_p,b_p)$ and two basis matrices $L$ and $R$
as:
\begin{eqnarray}
&&\varphi=L^{a_1}R^{b_1}L^{a_2}R^{b_2}\cdots L^{a_p}R^{b_p},\\
&&L=\left(
\begin{array}{cc}
1 & 1\\
0 & 1
\end{array}
\right), \quad
R=\left(
\begin{array}{cc}
1 & 0\\
1 & 1
\end{array}
\right).
\end{eqnarray}
For the simplest choice $\varphi=LR$, the manifold $T_{LR}$ 
is isomorphic to the figure eight knot complement.

The next simplest choice is $\varphi=L^2R$. 
The manifold $T_{L^2R}$ appears in the table of
the SnapPea census manifolds as $m009$ \cite{HW, magic, FKP}. 
$m009$ is also described as an arithmetic knot complement 
in $\mathbb{RP}^3$ \cite{BR}.\footnote{In \cite{Reid} it is shown that
the figure eight knot is the unique arithmetic knot in $\mathbb{S}^3$.}
$T_{L^2R}$ can be decomposed into three ideal tetrahedra.
The partition function of the state integral model is \cite{Hikami_gen}
\begin{equation}
Z_{\hbar}(T_{L^2R};\hbar,u)=\sqrt{2}e^{-u}\int d{\bf p}\delta_C({\bf p};u)
\langle p_1,p_5|{\bf S}^{-1}|p_6,p_3\rangle
\langle p_6,p_4|{\bf S}^{-1}|p_2,p_5\rangle
\langle p_3,p_2|{\bf S}|p_4,p_1\rangle.
\end{equation}
The shape parameters for each tetrahedra are
$z_1=e^{p_3-p_5}$, $z_2=e^{p_5-p_4}$ and $z_3=e^{p_1-p_2}$,
and the meridian condition is given by
\begin{eqnarray}
p_3-p_4-p_1+p_2=2u.
\end{eqnarray}
Integrating out the extra parameters, one obtains the partition function
for $T_{L^2R}$ with an incomplete structure:
\begin{eqnarray}
Z_{\hbar}(T_{L^2R};\hbar,u) &=&\frac{\sqrt{2}}{4\pi\hbar}\int_{{\cal
    C}}dp_1dp_2
\frac{\Phi_{\hbar}(-p_1-2u+i\pi+\hbar)}{\Phi_{\hbar}(-p_1-p_2-2u-i\pi-\hbar)\Phi_{\hbar}(2p_1+p_2+2u-i\pi-\hbar)}\nonumber
\\ &&\hspace*{2.3cm} \times
e^{-\frac{2}{\hbar}\left[u(u+p_1+p_2)+\frac{1}{2}(p_1+\frac{1}{2}p_2)^2-\frac{\pi^2}{12}-\frac{\hbar^2}{12}-\frac{\pi}{4}\hbar\right]-u}.
\end{eqnarray}
In this case the term $i\pi+\hbar$ in the argument of the quantum
dilogarithm functions cannot be removed by the shift of the parameters
as the figure eight knot case.

Expanding the integrand as above one obtains
\begin{eqnarray}
Z_{\hbar}(T_{L^2R};\hbar,u)&=&\frac{\sqrt{2}}{4\pi\hbar}
\int_{C}dp_1dp_2\;e^{\Upsilon(\hbar,p_1,p_2,u)},
\\
\Upsilon(\hbar,p_1,p_2,u)&=&\sum_{n=0}^{\infty}\Big[
B_n(\frac{1}{2}+\frac{-p_1-\hbar}{2\hbar})
{\rm Li}_{2-n}(\frac{1}{m^2x})
-B_n(\frac{1}{2}+\frac{-p_1-p_2+\hbar}{2\hbar})
{\rm Li}_{2-n}(\frac{1}{m^2xy})
\nonumber \\
&&\hspace{0.8cm}
-B_n(\frac{1}{2}+\frac{2p_1+p_2+\hbar}{2\hbar})
{\rm Li}_{2-n}(m^2x^2y)
\Big]\frac{(2\hbar)^{n-1}}{n!}
\nonumber \\
&&
-\frac{1}{2\hbar}\log (m^2)[\log (m^2x^2y^2)+2p_1+2p_2]
-\frac{1}{\hbar}\Big[\log(xy^{1/2})+p_1+\frac{1}{2}p_2\Big]^2
-u
\nonumber \\
&=&\sum_{k=-1}^{\infty}\sum_{i=0}^{\infty}\sum_{j=0}^{\infty}
\Upsilon_{i,j,k}(x,y,m)p_1^ip_2^j\hbar^k,
\end{eqnarray}
where $(x,y)=(e^{p_1},e^{p_2})$.
At the critical point $(e^{p_{10}},e^{p_{20}})=(x(u),y(u))$, 
the coefficients $\Upsilon_{1,0,-1}$ and $\Upsilon_{0,1,-1}$
vanishes. The solution for $\Upsilon_{1,0,-1}=0$ and 
$\Upsilon_{0,1,-1}=0$ which corresponds to the geometric branch is
\begin{eqnarray}
x(u)
&=&\frac{-1+m^2+m^4-\sqrt{1-2m^2-5m^4-2m^6+m^8}}{2(m^2+m^4)},\\
y(u)
&=&\frac{2m^2(1+2m^2+m^4-\sqrt{1-2m^2-5m^4-2m^6+m^8})}{(-1+m^2+m^4-\sqrt{1-2m^2-5m^4-2m^6+m^8})^2}.
\end{eqnarray}

Around the critical point, $Z_{\hbar}(T_{L^2R};\hbar,u)$ is expanded as:
\begin{eqnarray}
Z_{\hbar}(T_{L^2R};\hbar,u)
&=&\frac{e^{u+\frac{1}{\hbar}V(u)}}{2\sqrt{2}\pi\hbar}
\int_{C}dp_1dp_2 \;
e^{-\frac{b_{11}(u)p_1^2+b_{22}(u)p_2^2+b_{12}(u)p_1p_2}{2\hbar}}\nonumber \\
&&\hspace*{1cm}\times
\exp\bigg[
\frac{1}{\hbar}\sum_{i+j=3}^{\infty}\Upsilon_{i,j,-1}p_1^ip_2^j
+\sum_{i,j=0}^{\infty}\sum_{k=0}^{\infty}\Upsilon_{i,j,k}
p_1^ip_2^j\hbar^k
\bigg], \\
\label{exp_m009}
V(u)&=&{\rm Li}_2\Big(\frac{1}{m^2 x}\Big)-{\rm Li}_2\Big(\frac{1}{m^2
	       x y}\Big)-{\rm Li}_2(m^2x^2y)
\nonumber \\
&&-\log(m^2)\log(m^2x^2y^2)-2[\log(xy^{1/2})]^2+\frac{\pi^2}{6}.
\end{eqnarray}
From the equations (\ref{saddle}) and (\ref{NZ}) the A-polynomial
\cite{BRVD}:
\begin{eqnarray}
A_{T_{L^2R}}(l,m)=m^2+l(-1+2m^2+2m^4-m^6)+m^4l^2
\label{Aply_m009}
\end{eqnarray}
is found from the above potential $V(u)$.

In the geometric branch, the coefficients $b_{\alpha\beta}$
of the quadratic term yield
\begin{eqnarray}
b_{11}&=&\frac{1}{16m^2}\Big[
8 -9m^2 + 7 m^4 + m^6 
-(8+3m^2) \sqrt{1 - 2 m^2 - 5 m^4 - 2 m^6 + m^8} \Big], 
\nonumber \\
b_{22}&=&-\frac{1+m^2}{8m^2}\Big[
-1 +m^2 - m^4
- \sqrt{1 - 2 m^2 - 5 m^4 - 2 m^6 + m^8}
\Big],
\nonumber \\
b_{12}&=&-\frac{2+m^2+m^4+m^6-(2+m^2)\sqrt{1 - 2 m^2 - 5 m^4 - 2 m^6 + m^8}}{4m^2}.
\end{eqnarray}
The constant term $\Upsilon_{0,0,0}$ is
\begin{eqnarray}
e^{\Upsilon_{0,0,0}}
=\sqrt{\frac{(1+m^2+m^4-\sqrt{1 - 2 m^2 - 5 m^4 - 2 m^6 + m^8})}{2}}.
\end{eqnarray}
Then the 1-loop term $S_1(u)$ obeys
\begin{eqnarray}
\label{S1_m009}
&&\exp[S_1(u)]
=\frac{1}{\sqrt{2(b_{11}b_{22}-b_{12}^2/4)}}
e^{u+\Upsilon_{0,0,0}}
=\frac{im^2}{2}\Delta(m),\\
&&\Delta(m)=\sqrt{1 - 2 m^2 - 5 m^4 - 2 m^6 + m^8},
\end{eqnarray}
and this result coincides with the Reidemeister torsion \cite{Porti, DijFu}.

The higher order terms are obtained iteratively by expanding
(\ref{exp_m009}) and adopting a formula for the Gaussian integral
\begin{eqnarray}
\int f(\vec{x})e^{-\frac{1}{2}A_{ij}x^i x^j} d^nx=\sqrt{(2\pi)^n\over \det{A}}\left. \exp\left({1\over 2}(A^{-1})^{ij}{\partial \over \partial x^i}{\partial \over \partial x^j}\right)f(\vec{x})\right|_{\vec{x}=0}.
\end{eqnarray}
After some computations, one obtains the perturbative invariant $S_2(u)$:
\begin{eqnarray}
S_2(u)&=&\Upsilon_{0,0,1}
\nonumber \\
&&
+\frac{1}{b_0}
[b_{11}\Upsilon_{0,1,0}^2+2b_{11}^{(\alpha)}\Upsilon_{0,2,0}
+b_{22}\Upsilon_{1,0,0}^2+2b_{22}^{(\alpha)}\Upsilon_{2,0,0}
-b_{12}\Upsilon_{0,1,0}\Upsilon_{1,0,0}
-b_{12}\Upsilon_{1,1,0}]
\nonumber \\
&&+\frac{1}{b_0^2}
[12b_{11}^2(\Upsilon_{0,1,0}\Upsilon_{0,3,-1}+\Upsilon_{0,4,-1})
+12b_{22}^2(\Upsilon_{1,0,0}\Upsilon_{3,0,-1}+\Upsilon_{4,0,-1})
\nonumber \\
&&
-6b_{11}b_{12}
(\Upsilon_{0,3,-1}\Upsilon_{1,0,0}+\Upsilon_{0,3,-1}\Upsilon_{1,0,0}
+\Upsilon_{0,1,0}\Upsilon_{1,2,-1}+\Upsilon_{1,3,-1})
\nonumber \\
&&
-6b_{22}b_{12}
(\Upsilon_{3,0,-1}\Upsilon_{0,1,0}+\Upsilon_{3,0,-1}\Upsilon_{0,1,0}
+\Upsilon_{1,0,0}\Upsilon_{2,1,-1}+\Upsilon_{3,1,-1})
\nonumber \\ 
&&
+2(2b_{11}b_{22}+b_{12}^2)
(
\Upsilon_{1,0,0}\Upsilon_{1,2,-1}+\Upsilon_{0,1,0}\Upsilon_{2,1,-1}
+\Upsilon_{2,2,-1}
)]
\nonumber \\
&&
+\frac{1}{b_0^3}
[60b_{11}^3(\Upsilon_{0,3,-1}^2
-\Upsilon_{0,3,-1}\Upsilon_{1,2,-1})
+60b_{22}^3(\Upsilon_{3,0,-1}^2
-\Upsilon_{3,0,-1}\Upsilon_{2,1,-1})
\nonumber \\
&&
+12(b_{11}^2b_{22}+b_{11}b_{12}^2)
(\Upsilon_{1,2,-1}^2
+2\Upsilon_{0,3,-1}\Upsilon_{2,1,-1})
\nonumber \\
&&
+12(b_{22}^2b_{11}
+b_{22}b_{12}^2)
(\Upsilon_{2,1,-1}^2
+2\Upsilon_{3,0,-1}\Upsilon_{1,2,-1})
\nonumber \\
&&
+(36b_{11}b_{22}b_{12}+6b_{12}^3)
(\Upsilon_{1,2,-1}\Upsilon_{2,1,-1}
+\Upsilon_{0,3,-1}\Upsilon_{3,0,-1})
],
\end{eqnarray}
where $b_0=4b_{11}b_{22}-b_{12}^2$.
Plugging the explicit form of $\Upsilon_{i,j,k}$
in the geometric branch, one finds
\begin{equation}
S_2(u)=-\frac{5-11m^2+22m^4+105m^6+22m^8-11m^{10}+5m^{12}}
{48(1-2m^2-5m^4-2m^6+m^8)^{3/2}}+\frac{1}{16},
\label{S2_m009}
\end{equation}
where in this case $S_2(u=\pi i)$ has real and imaginary part.

The further higher order terms $S_3(u)$,  $S_4(u)$, and $S_5(u)$
are also computed in the same manner.
Plugging the explicit form of $\Upsilon_{i,j,k}(x,y,m)$ for the geometric
branch into this expansion, one finds
\begin{eqnarray}
&&S_3(u)=\frac{m^4(1-m^2+m^4)(1+9m^2+4m^4-9m^6+4m^8+9m^{10}+m^{12})}
{2(1-2m^2-5m^4-2m^6+m^8)^3},
\label{S3_m009}
\\
&&
S_4(u)=m^2 (1 - 68 m^2 - 3770 m^4 + 137 m^6 - 30073 m^8 - 58605 m^{10} + 
     104390 m^{12} + 20753 m^{14}
\nonumber \\
&&\quad\quad\quad\quad\quad
 - 222062 m^{16} + 20753 m^{18} + 
     104390 m^{20} - 58605 m^{22} - 30073 m^{24} + 137 m^{26} 
\nonumber \\
&&\quad\quad\quad\quad\quad
 - 3770 m^{28} - 68 m^{30} + m^{32})
/(720 (1 - 2 m^2 - 5 m^4 - 2 m^6 + m^8)^{9/2}),
\label{S4_m009}
\\
&&
S_5(u)=m^4 (1 + 86 m^2 + 179 m^4 + 3870 m^6 + 7447 m^8 - 7820 m^{10} + 
   51914 m^{12} + 60396 m^{14} 
\nonumber \\
&&\quad\quad\quad\quad\quad
- 183475 m^{16} - 25486 m^{18} 
   + 311325 m^{20} - 
   25486 m^{22} - 183475 m^{24} + 60396 m^{26} 
\nonumber \\
&&\quad\quad\quad\quad\quad
+ 51914 m^{28} - 7820 m^{30} + 
   7447 m^{32} + 3870 m^{34} + 179 m^{36} + 86 m^{38} + m^{40})
\nonumber \\
&&\quad\quad\quad\quad
/(24 (1 - 2 m^2 - 5 m^4 - 2 m^6 + m^8)^6).
\label{S5_m009}
\end{eqnarray}


\section{Free energy on character variety via topological recursion relations}

In \cite{EO} Eynard and Orantin defined a collection of symplectic
invariants ${\cal F}^{(g,0)}$, $g \in {\IN}\cup\{0\}$ for any complex
plane curve by means of a set of topological recursion relations. In
the context of matrix models, the complex plane curve is the spectral
curve, and the symplectic invariant ${\cal F}^{(g,0)}$ is the free
energy for genus $g$ \cite{Ambjorn,Akemann}. In this section using the
recursion relations, we define free energies ${\cal F}^{(g,h)}$ (for
genus $g$ with $h$ boundaries in ``world sheet language'') on the
character variety:
\begin{equation}
{\cal C}=\left\{x,l \in {\IC}^*~|~A(l,x)=0\right\}~\subset {\IC}^* \times {\IC}^*
\label{eqn:3.1}
\end{equation}
defined as the zero locus of the A-polynomial $A_K(l,m)$ reviewed in
section 2, where we redefined the parameters as
$A_K(l,m)=A(l,m^2)=A(l,x)$. The topological recursion relations iteratively
determine the free energies ${\cal F}^{(g,h)}$ (order by order in the
Euler number $\chi=2-2g-h$ in world sheet language). We compute the
free energies up to $\chi=-2$ for the two concrete examples described
in section 2.2 and 2.3, and compare the computation of the
perturbative CS expansion for them (see appendix D for the computation
of the free energy with $\chi=-3$).

\subsection{Eynard-Orantin topological recursion relation}

\begin{figure}[t]
 \begin{center}
  \includegraphics[width=140mm]{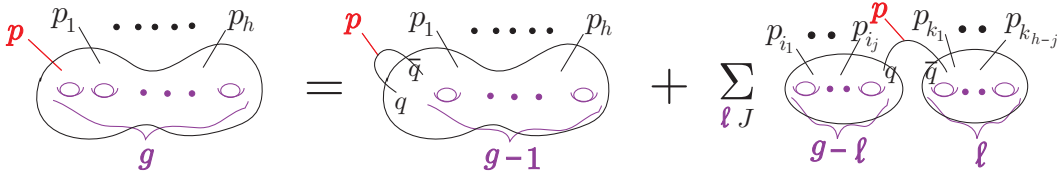}
 \end{center}
 \caption{Structure of the Eynard-Orantin topological recursion
   relation (\ref{eqn:3.2})}
 \label{fig:3.1}
\end{figure}

In this subsection we summarize the Eynard-Orantin topological
recursion relation, and its computation. Assuming that the branching
number at each ramification point $q_i$, $i=1,\ldots,n$ on the
character variety ${\cal C}$ is one, and then on neighborhood of
$q_i$, one finds two distinct points $q,{\bar q}\in {\cal C}$ such
that $x(q)=x(\bar{q})$ on the projected coordinate. The multilinear
meromorphic differentials $W^{(g,h)}(p_1,\ldots,p_h)$ on ${\cal C}$
are defined by the Eynard-Orantin topological recursion relation:
\begin{eqnarray}
&&W^{(0,1)}(p):=0,\quad
  W^{(0,2)}(p,q):=B(p,q), \hspace{19.5em}\nonumber\\ &&W^{(g,h+1)}(p,p_1,\ldots,p_h):=\sum_{q_i\in{\cal
      C}}\mathop{\mbox{Res}}_{q=q_i}
  \frac{dE_{q,\bar{q}}(p)}{\omega(q)-\omega(\bar{q})}\Bigl\{W^{(g-1,h+2)}(q,\bar{q},p_1,\ldots,p_h)
  \nonumber\\ &&\hspace{13em} +\sum_{\ell=0}^g \sum_{J \subset
    H}W^{(g-\ell,|J|+1)}(q,p_J)W^{(\ell,|H|-|J|+1)}(\bar{q},p_{H
    \backslash J})\Bigr\},
\label{eqn:3.2}
\end{eqnarray}
where $\omega(p)=\log l(p)dx(p)/x(p)$, and
$H=\{1,\ldots,h\},~J=\{i_1,\ldots,i_j\}\subset
H,~p_J=\{p_{i_1},\ldots,p_{i_j}\}$. The Bergman kernel $B(p,q)$,
which should be a planar two-point function of a chiral boson on ${\cal
C}$ as \cite{DVKS}, is defined by the conditions:
\begin{eqnarray}
&&\bullet~~B(p,q)\mathop{\sim}_{p \to q} \frac{dpdq}{(p-q)^2}+\mbox{finite}.\quad\quad \bullet~~\mbox{Holomorpic except}~p=q. \nonumber\\
&&\bullet~~\oint_{A_i}B(p,q)=0,\quad i=1,\ldots,{\overline g}=\mbox{the genus of ${\cal C}$},
\label{eqn:3.3}
\end{eqnarray}
where $A_i$ are the $A$-cycles in a canonical basis $(A_i,B^i)$ of
one-cycles on ${\cal C}$, and $dE_{q,\bar{q}}(p)$ is the third type
differential which is a one-form on $p$ and a multivalued function on
$q$ defined by the conditions:
\begin{equation}
\bullet~~dE_{q,\bar{q}}(p)\mathop{\sim}_{p \to q} -\frac{dp}{2(p-q)}+\mbox{finite}.\quad \bullet~~dE_{q,\bar{q}}(p)\mathop{\sim}_{p \to {\bar q}} \frac{dp}{2(p-q)}+\mbox{finite}. \quad
\bullet~~\oint_{A_i}dE_q(p)=0.
\label{eqn:3.4}
\end{equation}
The topological recursion relation (\ref{eqn:3.2}) is diagrammatically described as in Fig.\ref{fig:3.1}.

In the following we consider the case that the character variety
${\cal C}$ is a genus ${\overline g}$ (distinguished from the genus
$g$ of the world sheet) curve with two sheets. From the one-form
$\omega(p)=\log l(p)dx(p)/x(p)$, one defines
$y(p)dp=(\omega(p)-\omega({\bar p}))/2$ as proposed in \cite{Marino,
  BKMP}:
\begin{equation}
y(p)=M(p)\sqrt{\sigma(p)},\quad \sigma(p)=\prod_{i=1}^{2{\overline g}+2}(p-q_i),\quad M(p)=\frac{1}{p\sqrt{\sigma(p)}}\tanh^{-1}\Big[\frac{\sqrt{\sigma(p)}}{f(p)}\Big],
\label{eqn:3.5}
\end{equation}
where $\mbox{Re} (q_1) \leq \mbox{Re} (q_2) \leq \ldots \leq \mbox{Re}
(q_{2{\overline g}+2})$, $f(p)$ is a rational function in $p$, and
$M(p)$ is called the moment function in the context of matrix models
\cite{Ambjorn,Akemann}.  In this case in \cite{Eyn1} it is found that
the third type differential $dE_{q,\bar{q}}(p)$ has the form:
\begin{equation}
dE_{q,\bar{q}}(p)=-\frac{\sqrt{\sigma(q)}}{2\sqrt{\sigma(p)}}\Big(\frac{1}{p-q}-\sum_{i=1}^{{\overline g}}C_i(q)L_i(p)\Big)dp,\quad C_i(q)=\frac{1}{2\pi i}\oint_{q\not\in A_i}\frac{dp}{(p-q)\sqrt{\sigma(p)}},
\label{eqn:3.6}
\end{equation}
where we introduced the (normalized) basis of the holomorphic differentials $L_i(p)dp/\sqrt{\sigma(p)}$ on ${\cal C}$ by
\begin{equation}
\oint_{A_j}\frac{L_i(q)}{\sqrt{\sigma(q)}}dq=2\pi i\delta_{i,j},\quad L_i(q)=\sum_{j=1}^{{\overline g}}L_{j,i}q^{j-1},\quad i=1,\ldots,{\overline g}.
\label{eqn:3.7}
\end{equation}
Note that when $q$ approaches a branch point $q_i$, (\ref{eqn:3.6}) cannot be used, i.e. if a contour $A_i$ contains the point $q$, instead one must replace $C_j(q)$ with $C_j(q)+\delta_{j,i}/\sqrt{\sigma(q)}$. Using the relation
\begin{equation}
B(p,q)=dq\frac{\partial}{\partial q}\Big(\frac{dp}{2(p-q)}-dE_{q,\bar{q}}(p)\Big),
\label{eqn:3.8}
\end{equation}
between the Bergman kernel $B(p,q)$ and the third type differential $dE_{q,\bar{q}}(p)$, one finds that the Bergman kernel has the form:
\begin{eqnarray}
\label{eqn:3.9}
&&B(p,q)=\frac{dpdq}{\sqrt{\sigma(p)\sigma(q)}}\Big(\frac{\sqrt{\sigma(p)\sigma(q)}+F(p,q)}{2(p-q)^2}+\frac{H(p,q)}{4}\Big),\\
&&F(p,q):=\frac12\bigl(\sigma(p)+\sigma(q)\bigr)-\frac{p-q}{4}\bigl(\partial_p\sigma(p)-\partial_q\sigma(q)\bigr),
\label{eqn:3.10}
\end{eqnarray}
where $H(p,q)$ is a symmetric polynomial in $p$ and $q$ \cite{Eyn1}.

\begin{figure}[t]
 \begin{center}
  \includegraphics[width=150mm]{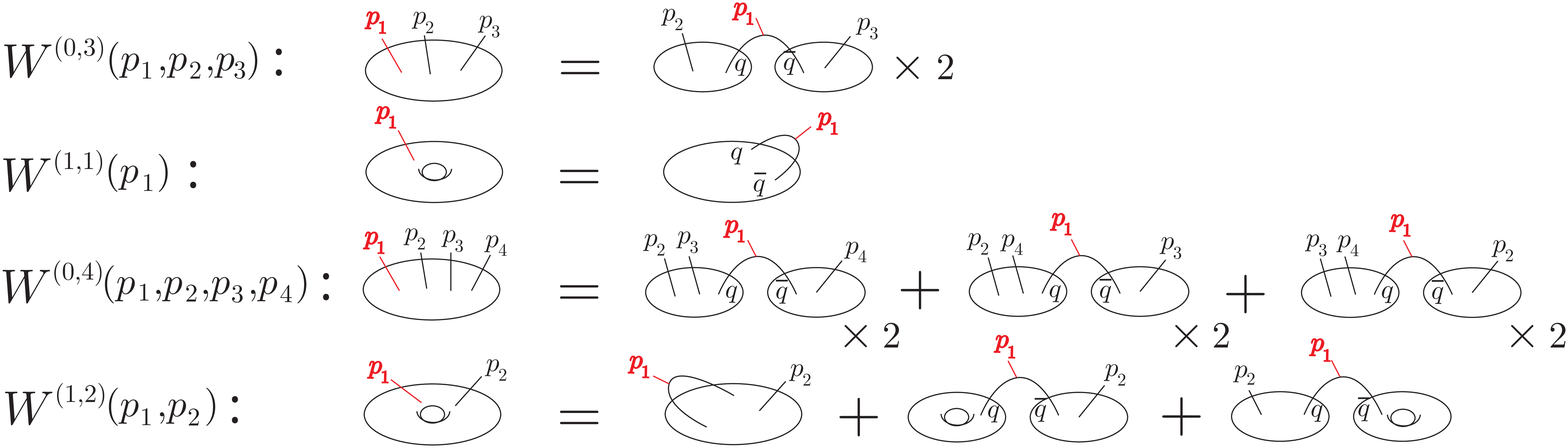}
 \end{center}
 \caption{Diagrammatic representation of (\ref{eqn:3.11}), (\ref{eqn:3.12}), (\ref{eqn:3.13}), and (\ref{eqn:3.14})}
 \label{fig:3.2}
\end{figure}

Let us compute the multilinear meromorphic differentials
$\widetilde{W}^{(g,h)}(p_1,\ldots,p_h)$ up to the Euler number $\chi=-2$
using the topological recursion relation (\ref{eqn:3.2}):
\begin{eqnarray}
\label{eqn:3.11}
&&W^{(0,3)}(p_1,p_2,p_3)=\sum_{q_i\in{\cal C}}\mathop{\mbox{Res}}_{q=q_i}\frac{2dE_{q,\bar{q}}(p_1)}{\omega(q)-\omega(\bar{q})}B(p_2,q)B(p_3,{\bar{q}}),\\
\label{eqn:3.12}
&&W^{(1,1)}(p_1)=\sum_{q_i\in{\cal C}}\mathop{\mbox{Res}}_{q=q_i}\frac{dE_{q,\bar{q}}(p_1)}{\omega(q)-\omega(\bar{q})}B(q,{\bar{q}}),\\
&&W^{(0,4)}(p_1,p_2,p_3,p_4)=\sum_{q_i\in{\cal C}}\mathop{\mbox{Res}}_{q=q_i}\frac{2dE_{q,\bar{q}}(p_1)}{\omega(q)-\omega(\bar{q})}\left\{B(p_2,\bar{q})W^{(0,3)}(p_3,p_4,q)+\mbox{perm}(p_2,p_3,p_4)\right\},\nonumber\\
&&
\label{eqn:3.13}
\\
\label{eqn:3.14}
&&W^{(1,2)}(p_1,p_2)=\sum_{q_i\in{\cal C}}\mathop{\mbox{Res}}_{q=q_i}\frac{dE_{q,\bar{q}}(p_1)}{\omega(q)-\omega(\bar{q})}\left\{W^{(0,3)}(p_2,q,\bar{q})+2W^{(1,1)}(q)B(p_2,{\bar{q}})\right\},
\end{eqnarray}
where these differentials are represented in Fig.\ref{fig:3.2}. One can expand these differentials by the kernel differentials \cite{BKMP},
\begin{eqnarray}
\chi_i^{(n)}(p):&=&\mathop{\mbox{Res}}_{q=q_i}\Big(-\frac{dE_{q,\bar{q}}(p)}{y(q)}\frac{1}{(q-q_i)^n}\Big) \nonumber\\
&=&\frac{dp}{2(n-1)!\sqrt{\sigma(p)}}\frac{\partial^{n-1}}{\partial q^{n-1}}\Big\arrowvert_{q=q_i}\frac{1}{M(q)}\Big(\frac{1}{p-q}-\sum_{i=1}^{{\overline g}}C_i(q)L_i(p)\Big),
\label{eqn:3.15}
\end{eqnarray}
where in the second equality (\ref{eqn:3.6}) is utilized. Using the relation (\ref{eqn:3.8}), one can expand $B(p,q)$ around $s^2=q-q_i=0$ as
\begin{eqnarray}
&&B(p,q)\simeq M_i\sqrt{\sigma_i'}ds \Bigl\{\chi_i^{(1)}(p)+3s^2\Bigl(\chi_i^{(2)}(p)+\Bigl(\frac{M_i'}{M_i}+\frac{\sigma_i''}{4\sigma_i'}\Bigr)\chi_i^{(1)}(p)\Bigr) \nonumber \\
&&+5s^4\Bigl(\chi_i^{(3)}(p)+\Bigl(\frac{M_i'}{M_i}+\frac{\sigma_i''}{4\sigma_i'}\Bigr)\chi_i^{(2)}(p)+\frac12\Bigl(\frac{M_i''}{M_i}+\frac{M_i'\sigma_i''}{2M_i\sigma_i'}+\frac{\sigma_i'''}{6\sigma_i'}-\frac{\sigma_i''^2}{16\sigma_i'^2}\Bigr)\chi_i^{(1)}(p)\Bigr)+{\cal{O}}(s^6)\Bigr\},\nonumber\\
&&\label{eqn:3.16}
\end{eqnarray}
where $M_i:=M(q_i),~\sigma_i':=\sigma'(q_i)$ etc., and the odd terms in
$s$ are ignored because these terms are irrelevant in the computation of
the topological recursion (\ref{eqn:3.2}). Therefore from (\ref{eqn:3.11}) one obtains \cite{BKMP},
\begin{equation}
W^{(0,3)}(p_1,p_2,p_3)=\frac12\sum_{i}M_i^2\sigma_i'\chi_i^{(1)}(p_1)\chi_i^{(1)}(p_2)\chi_i^{(1)}(p_3).
\label{eqn:3.17}
\end{equation}
When $q={\bar p}$ the Bergman kernel (\ref{eqn:3.9}) yields
\begin{equation}
B(p,{\bar p})=\lim_{q \to p}\frac{dpdp}{2(p-q)^2}\Big(1-\frac{F(p,q)}{\sqrt{\sigma(p)\sigma(q)}}\Big)-\frac{H(p)dpdp}{4\sigma(p)} =\frac{dpdp}{4}\Big(\frac{\sigma''(p)}{2\sigma(p)}-\frac{\sigma'(p)^2}{4\sigma(p)^2}-\frac{H(p)}{\sigma(p)}\Big),
\label{eqn:3.18}
\end{equation}
where $H(p):=H(p,p)$, and then this can be expanded around a branch point $p=q_i$ as:
\begin{equation}
B(p,{\bar p}) \simeq -\frac{dpdp}{4(p-q_i)}\Big\{\frac{1}{4(p-q_i)}+\Big(\frac{H(q_i)}{\sigma'(q_i)}-\frac{\sigma''(q_i)}{4\sigma'(q_i)}\Big)+{\cal O}(p-q_i)\Big\}.
\label{eqn:3.19}
\end{equation}
Using this expansion, from (\ref{eqn:3.12}) one finds \cite{BKMP}:
\begin{equation}
W^{(1,1)}(p_1)=\frac{1}{16}\sum_{i}\chi_i^{(2)}(p_1)+\frac{1}{4}\sum_{i}\Big(\frac{H(q_i)}{\sigma_i'}-\frac{\sigma_i''}{4\sigma_i'}\Big)\chi_i^{(1)}(p_1).
\label{eqn:3.20}
\end{equation}

To compute (\ref{eqn:3.13}) and (\ref{eqn:3.14}) let us write the kernel differentials in terms of the polynomials $F(p,q)$ and $H(p,q)$ by comparing the expansion (\ref{eqn:3.16}) with the expansion of (\ref{eqn:3.9}) around $s^2=q-q_i=0$,
\begin{eqnarray}
B(p,q)&=&\frac{dpdq}{4s \sqrt{\sigma(p)}}V(p,q;q_i), \quad V(p,q;q_i):=\frac{1}{\sqrt{\sigma(q;q_i)}}\Bigl(H(p,q)+\frac{2F(p,q)}{(p-q)^2}\Bigr),\hspace{1.6em} \nonumber\\
&\simeq&
\frac{dpdq}{4s \sqrt{\sigma(p)}}\Big\{V(p,q_i;q_i)+s^2 \partial_qV(p,q_i;q_i)+\frac{s^4}{2}\partial_q^2V(p,q_i;q_i)+{\cal{O}}(s^6)\Big\},
\label{eqn:3.21}
\end{eqnarray}
where $\sigma(q;q_i):=\sigma(q)/(q-q_i)$, and we have removed the term
$dpdq/2(p-q)^2$ in the expansion which is irrelevant in the computation
of the topological recursion (\ref{eqn:3.2}). Some of the kernel differentials are
\begin{eqnarray}
\label{eqn:3.22}
\chi_i^{(1)}(p)&=&\frac{dp}{2M_i\sigma_i'\sqrt{\sigma(p)}}{\widetilde V}(p,q_i),\quad {\widetilde V}(p,q):=H(p,q)+\frac{2F(p,q)}{(p-q)^2},\\
\label{eqn:3.23}
\chi_i^{(2)}(p)&=&\frac{dp}{6M_i\sigma_i'\sqrt{\sigma(p)}}\partial_q{\widetilde V}(p,q_i)-\Big(\frac{M_i'}{M_i}+\frac{\sigma_i''}{3\sigma_i'}\Big)\chi_i^{(1)}(p),\\
\chi_i^{(3)}(p)&=&\frac{dp}{20M_i\sigma_i'\sqrt{\sigma(p)}}\partial_q^2{\widetilde V}(p,q_i) \nonumber\\
\label{eqn:3.24}
&&\hspace{-1em}
-\Big(\frac{M_i''}{2M_i}+\frac{2M_i'\sigma_i''}{5M_i\sigma_i'}+\frac{\sigma_i'''}{10\sigma_i'}\Big)\chi_i^{(1)}(p)-\Big(\frac{M_i'}{M_i}+\frac{2\sigma_i''}{5\sigma_i'}\Big)\chi_i^{(2)}(p).
\end{eqnarray}

In the following for simplicity we only discuss the cases of ${\overline g}=1$. In this case the Bergman kernel is concretely given by the
Akemann's formula \cite{Akemann,Bouchard:2008gu}:
\begin{eqnarray}
\label{eqn:3.28}
&&
B(p,q)=\frac{dpdq}{\sqrt{\sigma(p)\sigma(q)}}\Big(\frac{\sqrt{\sigma(p)\sigma(q)}+f(p,q)}{2(p-q)^2}+\frac{G(k)}{4}\Big),\\
\label{eqn:3.29}
&&
f(p,q):=p^2q^2-\frac{1}{2}pq(p+q)S_1+\frac16(p^2+4pq+q^2)S_2-\frac12(p+q)S_3+S_4,\\
\label{eqn:3.30}
&&
G(k):=-\frac{1}{3}S_2+(q_1q_2+q_3q_4)-\frac{E(k)}{K(k)}(q_1-q_3)(q_2-q_4),\\
&&
K(k)=\int_0^1\frac{dt}{\sqrt{(1-t^2)(1-k^2t^2)}},~ E(k)=\int_0^1dt \sqrt{\frac{1-k^2t^2}{1-t^2}},
\label{eqn:3.27}
\end{eqnarray}
where $K(k)$, (resp. $E(k)$) is the complete elliptic integral of the first,
(resp. second) kind with the modulus $k^2=\frac{(q_1-q_2)(q_3-q_4)}{(q_1-q_3)(q_2-q_4)}$, and $S_k=\sum_{1\le j_1<j_2<\ldots<j_k \le 4}q_{j_1}q_{j_2}\cdots q_{j_k}$, $k=1,\ldots,4$ are the elementary symmetric polynomials of the branch points $q_i$.
By comparing (\ref{eqn:3.9}) with (\ref{eqn:3.28}), we see that ${\widetilde V}(p,q)$ in the kernel differentials is given by
\begin{equation}
{\widetilde V}(p,q)=G(k)+\frac{2f(p,q)}{(p-q)^2}.
\label{eqn:3.32}
\end{equation}
After some computation, (\ref{eqn:3.13}) and (\ref{eqn:3.14}) are expanded by the kernel differentials (see appendix B for the detailed derivation), and we summarize the results as follows:
\begin{eqnarray}
\label{eqn:3.33}
&&W^{(0,3)}(p_1,p_2,p_3)=\frac12\sum_{i}M_i^2\sigma_i'\chi_i^{(1)}(p_1)\chi_i^{(1)}(p_2)\chi_i^{(1)}(p_3),\\
\label{eqn:3.34}
&&W^{(1,1)}(p_1)=\frac{1}{16}\sum_{i}\chi_i^{(2)}(p_1)+\frac{1}{4}\sum_{i}\Big(\frac{G}{\sigma_i'}-\frac{\sigma_i''}{12\sigma_i'}\Big)\chi_i^{(1)}(p_1),\\
&&W^{(0,4)}(p_1,p_2,p_3,p_4)=
\frac14\sum_{i}\Big\{3M_i^2\Bigl(G+\frac23\sigma_i''+3\sigma_i'\frac{M_i'}{M_i}\Bigr)\chi_i^{(1)}(p_1)\chi_i^{(1)}(p_2)\chi_i^{(1)}(p_3)\chi_i^{(1)}(p_4) \nonumber\\
&&\quad~
+\sum_{j\neq i}M_iM_j\Bigl(G+\frac{2f(q_i,q_j)}{(q_i-q_j)^2}\Bigr)\Bigl(\chi_i^{(1)}(p_1)\chi_i^{(1)}(p_2)\chi_j^{(1)}(p_3)\chi_j^{(1)}(p_4)+\mbox{perm}(p_2,p_3,p_4)\Bigr) \nonumber\\
\label{eqn:3.35}
&&\quad~
+3M_i^2\sigma_i'\Bigl(\chi_i^{(1)}(p_1)\chi_i^{(1)}(p_2)\chi_i^{(1)}(p_3)\chi_i^{(2)}(p_4)+\mbox{perm}(p_1,p_2,p_3,p_4)\Bigr)\Big\},\\
&&
W^{(1,2)}(p_1,p_2)\nonumber\\
&&\quad~=
\frac{1}{32}\sum_{i}\Big\{\Big\{\frac{8G^2}{\sigma_i'^2}-\Bigl(\frac{2\sigma_i''}{3\sigma_i'^2}-\frac{11M_i'}{\sigma_i'M_i}\Bigr)G-\frac{\sigma_i''^2}{12\sigma_i'^2}-\frac{5\sigma_i'''}{18\sigma_i'}-\frac{7\sigma_i''M_i'}{6\sigma_i'M_i}+\frac{5M_i''}{2M_i}-\frac{3M_i'^2}{M_i^2} \nonumber\\
&&\quad~
+\sum_{j\neq i}\frac{M_i}{M_j\sigma_j'^2}\Big[\Bigl(4G-\frac23\sigma_j''-\sigma_j'\frac{M_j'}{M_j}\Bigr)\Bigl(G+\frac{2f(q_i,q_j)}{(q_i-q_j)^2}\Bigr)-\frac{\sigma_i'\sigma_j'}{3(q_i-q_j)^2}\Big]\Big\}\chi_i^{(1)}(p_1)\chi_i^{(1)}(p_2) \nonumber\\
&&\quad~
+\sum_{j\neq i}\frac{4}{\sigma_i'\sigma_j'}\Big(G+\frac{2f(q_i,q_j)}{(q_i-q_j)^2}\Big)^2\chi_i^{(1)}(p_1)\chi_j^{(1)}(p_2)+5\Big(\chi_i^{(1)}(p_1)\chi_i^{(3)}(p_2)+(p_1\leftrightarrow p_2)\Big)\nonumber\\
&&\quad~
+\Bigl(\frac{12G}{\sigma_i'}-\frac{\sigma_i''}{2\sigma_i'}+\frac{2M_i'}{M_i}\Bigr)\left(\chi_i^{(1)}(p_1)\chi_i^{(2)}(p_2)+(p_1\leftrightarrow p_2)\right)+3\chi_i^{(2)}(p_1)\chi_i^{(2)}(p_2)\Big\}.
\label{eqn:3.36}
\end{eqnarray}
In the following, we define free energies on character variety, and compute them using the above results for two concrete examples described in section 2.2 and 2.3 where the character varieties are reduced to genus one curve with two sheets on the geometric branch.

\subsection{Free energy on character variety}

Let us concentrate on the genus one case in (\ref{eqn:3.5}):
\begin{eqnarray}
&&y(p)=M(p)\sqrt{\sigma(p)},\quad \sigma(p)=\prod_{i=1}^4(p-q_i)=p^4-S_1p^3+S_2p^2-S_3p+S_4,\nonumber\\
&&M(p)=\frac{1}{p\sqrt{\sigma(p)}}\tanh^{-1}\Big[\frac{\sqrt{\sigma(p)}}{f(p)}\Big]=\frac{1}{2p\sqrt{\sigma(p)}}\log\frac{f(p)+\sqrt{\sigma(p)}}{f(p)-\sqrt{\sigma(p)}}.
\label{eqn:3.37}
\end{eqnarray}
In the following by taking the reciprocality of the A-polynomial into consideration, we assume that
\begin{equation}
S_1=S_3,\quad S_4=1,\quad q_2=q_1^{-1},\quad q_4=q_3^{-1}.
\label{eqn:3.38}
\end{equation}

The free energy $\widetilde{\cal F}^{(g,h)}(p_1,\ldots,p_h)$ is defined by 
\cite{Marino,BKMP}:
\begin{eqnarray}
\widetilde{\cal F}^{(g,h)}(p_1,\ldots,p_h)=\int^{p_1}\cdots\int^{p_h}
W^{(g,h)}(p_1,\ldots,p_h).
\end{eqnarray}
This free energy is related to the open topological string amplitude 
$F(V)$ in the A-model on the local toric Calabi-Yau 3-fold $X$
whose mirror curve is the spectral curve in the Eynard-Orantin
topological recursion relation \cite{Marino}:
\begin{eqnarray}
&&F(V)=\log Z(V),\quad 
Z(V)=\frac{1}{Z_{\rm closed}(X)}\sum_R Z_R(X){\rm Tr}_R V,
\\
&&V={\rm diag}(\xi_1,\xi_2,\ldots,\xi_n),
\end{eqnarray}
where $\xi_i$ ($i=1,\ldots,n$) denotes the location of the D-branes 
in the toric Calabi-Yau 3-fold $X$.
\begin{figure}[t]
 \begin{center}
  \includegraphics[width=70mm]{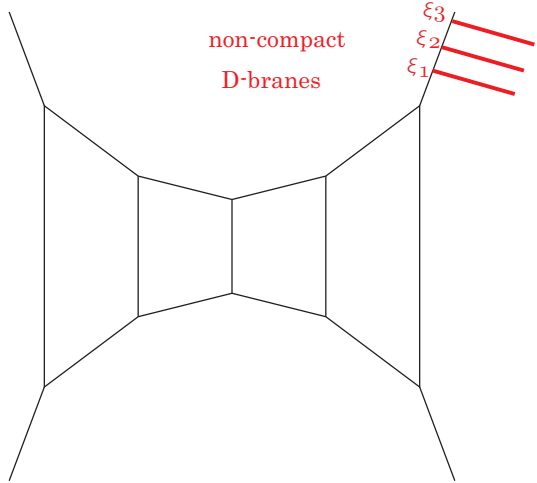}
 \end{center}
 \caption{Toric branes inserted on the local Calabi-Yau $3$-fold}
 \label{fig:toric}
\end{figure}
$Z_R(X)$ is the topological vertex amplitude on $X$ where the representation $R$
is assigned to the external leg that the D-brane is inserted.

The functional $F(V)$ is expanded as follows:
\begin{eqnarray}
F(V)&=&\sum_{g=0}^{\infty}\sum_{h=1}^{\infty}\sum_{w_1,\ldots,w_h}\frac{1}{h!}g_s^{2g-2+h}F_{\{w\},g}
{\rm Tr}V^{w_1}\cdots{\rm Tr}V^{w_h}.
\end{eqnarray}
After identifying
\begin{eqnarray}
{\rm Tr}V^{w_1}\cdots{\rm Tr}V^{w_h}\quad
\longleftrightarrow\quad
p_1^{w_1}\cdots p_h^{w_h},
\end{eqnarray}
the functional $F(V)$ is related to 
the free energy $\widetilde{\cal F}^{(g,h)}$:
\begin{eqnarray}
F(V)\quad
\longleftrightarrow\quad
\sum_{g=0}^{\infty}\sum_{h=1}^{\infty}g_s^{2g-2+h}\widetilde{\cal
 F}^{(g,h)}(p_1,\ldots, p_h).
\end{eqnarray}

As an assumption of our proposal, we introduce the D-branes 
whose locations are specified formally by
\begin{eqnarray}
V=\left(
\begin{array}{cc}
p & 0 \\
0 & p^{-1}
\end{array}
\right).
\end{eqnarray}
This choice of $V$ is nothing but the $PSL(2;\mathbb{C})$ holonomy representation matrix $\rho(\mu)$
along the meridian cycle $\mu$, and 
this free energy respects the reciprocality of the character variety as
was considered in \cite{DijFu}. Actually we see that the free energy ${\cal F}(p)$ is obtained from the free energy $\widetilde{\cal F}(p)$ for one brane at a point $p$ by ${\cal F}(p)=\widetilde{\cal F}(p+p^{-1})$.  Because the multilinear meromorphic differentials $W^{(g,h)}(p_1,\ldots,p_h)$ can be expanded in terms of the kernel differentials (\ref{eqn:3.15}) as in (\ref{eqn:3.33}) - (\ref{eqn:3.36}), it is convenient to introduce averaged kernel differentials:
\begin{eqnarray}
\widetilde{\chi}_i^{(n)}(p):&=&\chi_i^{(n)}(p)+\chi_i^{(n)}(p^{-1})=\frac{dp}{2(n-1)!\sqrt{\sigma(p)}}\frac{\partial^{n-1}}{\partial q^{n-1}}\Big\arrowvert_{q=q_i}\frac{1}{M(q)}\Big(\frac{1}{p-q}-\frac{1}{p^{-1}-q}\Big) \nonumber\\
&=&\frac{dw}{2(n-1)!\sqrt{\widetilde{\sigma}(w)}}\frac{\partial^{n-1}}{\partial q^{n-1}}\Big\arrowvert_{q=q_i}\frac{1}{qM(q)}\frac{1}{w-\alpha},
\label{eqn:3.40}
\end{eqnarray}
where $w=(p+p^{-1})/2$, $\alpha=(q+q^{-1})/2$. We have defined
\begin{equation}
\widetilde{\sigma}(w):=\frac{\sigma(p)}{p^2}=4w^2-2S_1w+(S_2-2)=4(w-\alpha_1)(w-\alpha_2),
\label{eqn:3.41}
\end{equation}
where $\alpha_1=(q_1+q_1^{-1})/2=(q_2+q_2^{-1})/2$ and $\alpha_2=(q_3+q_3^{-1})/2=(q_4+q_4^{-1})/2$. 
Using (\ref{eqn:3.40}) by replacing $\chi_i^{(n)}(p)$ with $\widetilde{\chi}_i^{(n)}(p)$, we define averaged multilinear meromorphic differentials $\widetilde{W}^{(g,h)}(p_1,\ldots,p_h)$ for $(g,h)\neq (0,1),(0,2)$, and define free energies ${\cal F}^{(g,h)}(p)$ (according to remodeling the B-model \cite{Marino, BKMP}) for the two toric branes on the character variety ${\cal C}$ as:
\begin{eqnarray}
\label{eqn:3.45}
&&{\cal F}^{(g,h)}(p):=\frac{1}{h!}{\cal F}^{(g,h)}(p,\ldots,p),\\
\label{eqn:3.46}
&&{\cal F}^{(g,h)}(p_1,\ldots,p_h):=\int^{p_1}\int^{p_2}\cdots \int^{p_h} {\overline W}^{(g,h)}(p_1',\ldots,p_h'),\\
\label{eqn:3.47}
&&{\overline W}^{(0,1)}(p):=\omega(p)+\omega(p^{-1}):=\log l(p)\frac{dp}{p}+\log l(p^{-1})\frac{dp^{-1}}{p^{-1}},\\
\label{eqn:3.48}
&&{\overline W}^{(0,2)}(p_1,p_2):=2B(p_1,p_2)+2B(p_1,p_2^{-1})-\frac{2dw_1dw_2}{(w_1-w_2)^2},\quad w_i=\frac{p_i+p_i^{-1}}{2},\\
\label{eqn:3.49}
&&{\overline W}^{(g,h)}(p_1,\ldots,p_h):=\widetilde{W}^{(g,h)}(p_1,\ldots,p_h)~ \mbox{for}~(g,h)\neq (0,1),(0,2),
\end{eqnarray}
where the factor $h!$ in (\ref{eqn:3.45}) is the symmetric factor. In (\ref{eqn:3.48}) the factor $2$ comes from $B(p^{-1}_1,p^{-1}_2)+B(p^{-1}_1,p_2)$ where $B(p_1,p_2)=B(p^{-1}_1,p^{-1}_2)$, and the term $dw_1dw_2/(w_1-w_2)^2$ needs for the regularization (exclusion of the double pole) of the Bergman kernel at $p_1=p_2$. By introducing a coupling constant $g_s$, we define
\begin{equation}
{\cal F}(p):=\frac12 \sum_{g=0,h=1}^{\infty}g_s^{2g-2+h}{\cal F}^{(g,h)}(p)=\sum_{n=0}^{\infty}{\tilde \hbar}^{n-1}{\cal F}_{n}(p)
\label{eqn:3.50}
\end{equation}
on the character variety ${\cal C}$, where to express ``chiral part''
of the free energy we insist the necessity of the factor
$1/2$.\footnote{ The meaning of ``chiral part'' may come from
  $SL(2;\mathbb{C})$ Chern-Simons gauge theory. The partition function
  of $SL(2;\mathbb{C})$ Chern-Simons gauge theory is
  holomorphically factorized as
  $Z^{SL(2;\mathbb{C})}(M;t,\bar{t})=Z(M;t)\bar{Z}(M;\bar{t})$ where
  $t, \bar{t}$ are coupling constants.  The factor $1/2$ would be
  interpreted as the holomorphic factorization.  } We also introduced
a new coupling constant ${\tilde \hbar}=g_s/2$ for a consistency with
the coupling constant $\hbar$ in the Chern-Simons gauge theory.

Note that in the definition (\ref{eqn:3.46}), there are ambiguities of
the integration constants. In this paper we claim that, by taking the
universal part which does not depend on the choice of the integration
constants, we obtain
\begin{equation}
{\cal F}_{n}(p) \simeq S_n^{({\rm geom})}(u),\quad p=m^2=e^{2u}.
\label{eqn:3.51}
\end{equation}
Here $S_n^{({\rm geom})}(u)$ is the perturbative invariant on the
geometric branch discussed in section 2. In this claim we neglect the
constant term in $S_n^{({\rm geom})}(u)$ which does not depend on $u$.
In the left hand side of the above claim, regularizations of $G(k)^n$
in (\ref{eqn:3.30}), as explained in the following, are needed.

In the rest of this section, to check this claim we compute
\begin{eqnarray}
\label{eqn:3.52}
&&{\cal F}_2(p)={\cal F}^{(0,3)}(p)+{\cal F}^{(1,1)}(p),\\
\label{eqn:3.53}
&&{\cal F}_3(p)=2{\cal F}^{(0,4)}(p)+2{\cal F}^{(1,2)}(p),
\end{eqnarray}
for the two examples of section 2.2 and 2.3, and find that the
different regularizations for $G(k)$ in the Bergman Kernel
(\ref{eqn:3.28}), and its square $G(k)^2$ are needed as:
\begin{eqnarray}
\label{eqn:3.54}
&&G(k)=-\frac{1}{3}S_2+2-\frac{E(k)}{K(k)}(q_1-q_3)(q_2-q_4)~\Rightarrow~G_1:=-\frac{1}{3}S_2+2,\hspace{5em}\\
\label{eqn:3.55}
&&G(k)^2~\Rightarrow~G_2:=G_1^2-(1-k^2)(q_1-q_3)^2(q_2-q_4)^2=G_1^2-\big(S_1^2-4(S_2-2)\big),
\end{eqnarray}
for the examples, where $k$ is the modulus of the elliptic integrals
$K(k)$ and $E(k)$ defined in (\ref{eqn:3.27}). 
The constant $G$ is determined uniquely by imposing zero A-period condition. 
But we have to impose some ad-hoc regularizations to $G^n$ terms in the
free energy. This regularization may be compensating the some subtleties 
of the correspondence in the higher order terms of $\hbar$ expansion.
The subtleties may come from the choice of the integration contour for
the BKMP's free energy or ${\cal O}(\hbar/2)$ shift of the moduli 
of the A-polynomial.\footnote{The similar problem occurs in the inner
toric brane computation to realize the $2D/4D$ instanton partition function
of the four dimensional ${\cal N}=2$ gauge theory in the AGT context \cite{AFKMY}.}
Although we do not know the general rule for this regularizations,  we 
heuristically find the rules which are applicable to some lower order
terms in the WKB expansion.
In appendix {\ref{sectionF} we compute the free energy ${\cal F}_4(p)$, and for the figure eight knot complement we find that different regularizations for $G(k)$ in ${\cal F}^{(1,3)}(p)$ and in ${\cal F}^{(2,1)}(p)$ are needed as in (\ref{eqn:F13}).

The leading term in the correspondence (\ref{eqn:3.51}) is understood as follows. By the reciprocality of the A-polynomial, the character variety has the property $l(p)l(p^{-1})=1$. Therefore we obtain
\begin{equation}
{\cal F}_0(p)=\frac{1}{4}{\cal F}^{(0,1)}(p)=\frac{1}{2}\int^p \log l(p)\frac{dp}{p}=\int^m \log l(m)\frac{dm}{m},
\label{eqn:3.56}
\end{equation}
and this is nothing but $S_0^{({\rm geom})}(u)$, where $p=m^2$, except a constant shift \cite{NZ,Yoshida}. The subleading term ${\cal F}_1(p)={\cal F}^{(0,2)}(p)/2$ is discussed in \cite{DijFu} (see also appendix C). In the rest of this section we check the above claim for two concrete examples in section 2.

\subsection{Figure eight knot complement}

As the first example, from the A-polynomial (\ref{AplyFig}) of the figure eight knot, we obtain the data of the curve:
\begin{eqnarray}
\label{eqn:3.57}
&&\sigma(p)=p^4-2p^3-p^2-2p+1,\quad \widetilde{\sigma}(w)=4w^2-4w-3,\\
\label{eqn:3.58}
&&f(p)=\frac{p^4-p^3-2p^2-p+1}{p^2-1},
\end{eqnarray}
where $w=(p+p^{-1})/2$, and $\widetilde{\sigma}(w)=\sigma(p)/p^2$.

The free energy ${\cal F}_2(p)$ defined in (\ref{eqn:3.52}) is computed from (\ref{eqn:3.33}) and (\ref{eqn:3.34}):
\begin{eqnarray}
&&
\frac{\overline{W}^{(0,3)}(p_1,p_2,p_3)}{dw_1dw_2dw_3}=\frac{8\big[(2w_1-3)(2w_2-3)(2w_3-3)+(2w_1+1)(2w_2+1)(2w_3+1)\big]}{\widetilde{\sigma}(w_1)^{3/2}\widetilde{\sigma}(w_2)^{3/2}\widetilde{\sigma}(w_3)^{3/2}},\nonumber\\
\label{eqn:3.59}
&&\\
&&
\frac{\overline{W}^{(1,1)}(p)}{dw}=\frac{-(2w-3)^2+(2w+1)^2}{4\widetilde{\sigma}(w)^{5/2}}+\frac{-5(3G+2)(2w-3)+3(3G-2)(2w+1)}{180\widetilde{\sigma}(w)^{3/2}}.\nonumber\\
\label{eqn:3.60}
&&
\end{eqnarray}
Thus from (\ref{eqn:3.45}) and (\ref{eqn:3.52}) we obtain
\begin{eqnarray}
\label{eqn:3.61}
&&
{\cal F}^{(0,3)}(p)=-\frac{12w^2-12w+7}{12\widetilde{\sigma}(w)^{3/2}},\\
\label{eqn:3.62}
&&
{\cal F}^{(1,1)}(p)=-\frac{1}{3\widetilde{\sigma}(w)^{3/2}}-\frac{2(6G+1)w-9(G+1)}{180\widetilde{\sigma}(w)^{1/2}},\\
\label{eqn:3.63}
&&
{\cal F}_2(p)=-\frac{1}{12\widetilde{\sigma}(w)^{3/2}}\Big\{\frac{8(6G+1)}{15}w^3-\frac{4(21G-34)}{15}w^2-10w+\frac{9G+64}{5}\Big\}.
\end{eqnarray}
Using the regularization (\ref{eqn:3.54}), by replacing $G$ with $G_1=7/3$ we get
\begin{equation}
{\cal F}_2(p)=-\frac{1}{12\widetilde{\sigma}(w)^{3/2}}(8w^3-4w^2-10w+17).
\label{eqn:3.64}
\end{equation}
This coincides with the perturbative invariant (\ref{S2Fig}) by identifying the parameter $w=(m^2+m^{-2})/2$.

Next from (\ref{eqn:3.35}) and (\ref{eqn:3.36}) we compute the free energy ${\cal F}_3(p)$ defined in (\ref{eqn:3.53}).
As the result by (\ref{eqn:3.45}) and (\ref{eqn:3.53}) we obtain
\begin{eqnarray}
\label{eqn:3.67}
&&
{\cal F}^{(0,4)}(p)=\frac{1}{\widetilde{\sigma}(w)^3}\Big(\frac43w^5-\frac83w^4+2w^3+\frac{11}{3}w^2-\frac{67}{12}w+\frac{25}{12}\Big),\hspace{7.5em}\\
\label{eqn:3.68}
&&
{\cal F}^{(1,2)}(p)=\frac{1}{\widetilde{\sigma}(w)^3}\Big(\frac{16}{81}w^6-\frac{568}{405}w^5+\frac{821}{405}w^4+\frac{154}{45}w^3-\frac{83}{30}w^2-\frac{61}{15}w+\frac{859}{240}\Big) \nonumber\\
&&\hspace{5em}
+\frac{G}{\widetilde{\sigma}(w)^2}\Big(\frac{8}{135}w^4-\frac{29}{135}w^3+\frac{13}{90}w^2+\frac{7}{20}w-\frac{9}{40}\Big)+\frac{G^2(4w-3)^2}{3600\widetilde{\sigma}(w)},\\
\label{eqn:3.69}
&&
{\cal F}_3(p)=\frac{2}{\widetilde{\sigma}(w)^3}\Big(\frac{16}{81}w^6-\frac{28}{405}w^5-\frac{259}{405}w^4+\frac{244}{45}w^3+\frac{9}{10}w^2-\frac{193}{20}w+\frac{453}{80}\Big) \nonumber\\
&&\hspace{5em}
+\frac{2G}{\widetilde{\sigma}(w)^2}\Big(\frac{8}{135}w^4-\frac{29}{135}w^3+\frac{13}{90}w^2+\frac{7}{20}w-\frac{9}{40}\Big)+\frac{G^2(4w-3)^2}{1800\widetilde{\sigma}(w)},
\end{eqnarray}
and using the regularization (\ref{eqn:3.54}) and (\ref{eqn:3.55}), by replacing $G$ with $G_1=7/3$ and $G^2$ with $G_2=-95/9$ we get
\begin{equation}
{\cal F}_3(p)=\frac{2}{\widetilde{\sigma}(w)^3}(8w^3-4w^2-10w+7).
\label{eqn:3.70}
\end{equation}
This also coincides with the perturbative invariant (\ref{S3Fig}).

\subsection{Once punctured torus bundle over $\mathbb{S}^1$ with holonomy $L^2 R$}

As the second example, from the A-polynomial (\ref{Aply_m009}) of the once punctured torus bundle over $\mathbb{S}^1$ with holonomy $L^2 R$, we obtain the data of the curve:
\begin{eqnarray}
\label{eqn:3.71}
&&\sigma(p)=p^4-2p^3-5p^2-2p+1,\quad \widetilde{\sigma}(w)=4w^2-4w-7,\\
\label{eqn:3.72}
&&f(p)=\frac{p^3-2p^2-2p+1}{p-1},
\end{eqnarray}
where $\widetilde{\sigma}(w)=\sigma(p)/p^2$.

As same as the computation in section 3.3, the free energies ${\cal F}^{(0,3)}(p)$ and ${\cal F}^{(1,1)}(p)$ are obtained as:
\begin{eqnarray}
\label{eqn:3.73}
&&{\cal F}^{(0,3)}(p)=-\frac{8w^3+36w^2+6w+19}{48\widetilde{\sigma}(w)^{3/2}},\hspace{18em}\\
\label{eqn:3.74}
&&{\cal F}^{(1,1)}(p)=-\frac{(72G-40)w^3-(156G-12)w^2-(42G-210)w+147G+217}{336\widetilde{\sigma}(w)^{3/2}},
\end{eqnarray}
and using the regularization (\ref{eqn:3.54}), by replacing $G$ with $G_1=11/3$ we find
\begin{equation}
{\cal F}_2(p)=-\frac{1}{48\widetilde{\sigma}(w)^{3/2}}(40w^3-44w^2+14w+127).
\label{eqn:3.75}
\end{equation}
This coincides with the perturbative invariant (\ref{S2_m009}) by identifying the parameter $w=(m^2+m^{-2})/2$.

The free energies ${\cal F}^{(0,4)}(p)$ and ${\cal F}^{(1,2)}(p)$ are also computed. Using the regularization (\ref{eqn:3.54}) and (\ref{eqn:3.55}), by replacing $G$ with $G_1=11/3$ and $G^2$ with $G_2=-167/9$ we find
\begin{equation}
{\cal F}_3(p)=-\frac{1}{128\widetilde{\sigma}(w)^3}(64w^6-192w^5-1168w^4-3488w^3+2300w^2+2996w-2071).
\label{eqn:3.76}
\end{equation}
For the free energy ${\cal F}_3(p)$ we should consider an imaginary term corresponding to the Chern-Simons term of the partition function for the state integral model as in (\ref{S2_m009}). Such the contribution, if we add a constant term $1/128$:
\begin{equation}
{\cal F}_3(p)+\frac{1}{128}=\frac{1}{2\widetilde{\sigma}(w)^3}(16w^4+64w^3-32w^2-56w+27),
\label{eqn:3.77}
\end{equation}
then this coincides with the perturbative invariant (\ref{S3_m009}).

\section{Torus knots}
\baselineskip 17.5pt
In this section, we will further discuss the correspondence for 
torus knots. 
The perturbative invariants $S_k(u)$
and the BKMP's free energies ${\cal F}_k$ are computed exactly for this
case.
Although the results are rather trivial on both sides, we are able to
check the correspondence exactly for this example.

\subsection{Colored Jones polynomial for torus knots}
A torus knot is described as a curve on a two-torus $\mathbb{T}^2$, 
and a pair of coprime integers $(p,q)$ specifies the number of windings around 
each cycle of $\mathbb{T}^2$. For the $(p,q)$ torus knot
the colored Jones polynomial is found explicitly \cite{Mor}.

Although a torus knot does not admit a hyperbolic structure on the
$\mathbb{S}^3$ complement, the asymptotic behavior of the 
the colored Jones polynomials is studied in the context of 
the Melvin-Morton-Rozansky conjecture \cite{MMR,Mor,Roz1,GL,Roz2} and 
the volume conjecture \cite{KT,DK,HM2}.
In the analysis of the volume conjecture, the volume of the torus knot
complement vanishes but the Chern-Simons invariant \cite{KK} is realized
as the asymptotic limit of the colored Jones polynomial around 
the exponential growth point.

Furthermore, the $q$-difference equation for the $(p,q)$ torus knots have been 
found explicitly \cite{Garou1,Garou2,Garou-Le,Geronimo,HikamiAJ}.
Adopting the technique of \cite{DGLZ}, we will perform the WKB expansion
iteratively from the $q$-difference equation.

\subsection{AJ conjecture for torus knots}
The $q$-difference equations for $(2,2\mathfrak{m}+1)$ torus knots $T_{2,2\mathfrak{m}+1}$ 
are found from the inhomogeneous difference equation 
\cite{HikamiAJ}:
\begin{eqnarray}
&&J_n({T_{2,2\mathfrak{m}+1}};q)=q^{\mathfrak{m}(n-1)}\frac{1-q^{2n-1}}{1-q^n}-q^{(2\mathfrak{m}+1)n-\mathfrak{m}}
\frac{1-q^{n-1}}{1-q^n} J_{n-1}({T_{2,2\mathfrak{m}+1}};q),
\label{inhom_2_2m+1}
\\
&&J_n({T_{s,t}};q)=\frac{q^{\frac{1}{2}(s-1)(t-1)(n-1)}}{1-q^n}
(1-q^{s(n-1)+1}-q^{t(n-1)+1}+q^{(s+t)(n-1)})
\nonumber \\
&&\quad\quad\quad\quad\quad\;\;
+\frac{1-q^{n-2}}{1-q^n}q^{st(n-1)+1}J_{n-2}(T_{s,t};q).
\end{eqnarray}
As was firstly calculated in \cite{Garou1,Garou2}, 
one can obtain the homogeneous $q$-difference equation
from the inhomogeneous one by adopting {\tt Mathematica} 
packages `{\tt qZeil.m}' and `{\tt qMultiSum.m}' developed by Paule and Riese
\cite{qZeil}.

For the trefoil knot ${\bf 3_1}$ the $q$-difference equation for the
colored Jones polynomial is \cite{Geronimo}:
\begin{eqnarray}
&& P_{\bf 3_1}(E,Q)J_{n}({\bf 3_1};q)=0, \\
&& P_{\bf 3_1}(E,Q)=\frac{q^2 Q^2(q^2-q^2 Q)}{q^3-q^4 Q^2}
\nonumber \\
&&\quad\quad\quad\quad\quad\quad
+\frac{(q-q^2 Q)(q+q^2 Q)(q^4-q^5 Q+q^6 Q^2-q^7 Q^2-q^7 Q^3+q^8
Q^4)}{q^2Q(q-q^4Q^2)(q^3-q^4Q^2)}E
\nonumber \\
&&\quad\quad\quad\quad\quad\quad
+\frac{-1+q^2Q}{Q(q-q^4Q^2)}E^2,
\end{eqnarray}
where 
\begin{eqnarray}
 (Qf)(n)=q^nf(n), \quad (Ef)(n)=f(n+1), \quad
q=e^{2\hbar}.
\end{eqnarray}
In $\hbar\to 0$ limit, the polynomial $P_{\bf 3_1}$ yields
\begin{eqnarray}
P_{\bf 3_1}(L,M)=-\frac{(L-1)(L+M^3)}{M(1+M)}.
\end{eqnarray}
The numerator of $P_{\bf 3_1}(L,M)$ is the A-polynomial for the trefoil knot.

The expectation value of the Wilson loop operator $W_{n}(K;q)$ is
different from the colored Jones polynomial $J_n(K;q)$ by the unknot
factor $J_{n}({\rm unknot};q)$.
By factoring out $q^{j/2}Q-q^{-j/2}$
in the coefficient of $E^j$,
one obtains the $q$-difference equation for $W_{n}(K;q)$.
The Wilson loop expectation value 
$W_{n}(K;q)$ is identified with $Z(S^3\backslash K;\hbar,u)$
for $q=e^{2\hbar}$ and $m=q^{n/2}$, if $K$ is the hyperbolic knot.

We rewrite the $q$-difference equation for $W_{n}({\bf 3_1};q)$
following the notation of \cite{DGLZ} as $Q=\hat{m}^2$:
\begin{eqnarray}
&& \sum_{j=0}^2 a_j(q^{n/2};q)W_{n+j}({\bf 3_1};q)=0,
\\
&&a_0(\hat{m},q)=\frac{q^2 \hat{m}^4}{q\hat{m}^4-1}, 
\nonumber \\
&&a_1(\hat{m},q)=-\frac{q^{1/2}(1+q\hat{m}^2)[1-q\hat{m}^2-(q^3-q^2)\hat{m}^4-q^3\hat{m}^6+q^4\hat{m}^8]}{\hat{m}^2(1-q^3\hat{m}^4)(1-q\hat{m}^4)},
\nonumber \\
&&a_2(\hat{m},q)=\frac{1}{\hat{m}^2(1-q^3\hat{m}^4)}.
\nonumber 
\end{eqnarray}
Furthermore $q$-difference operator $P_{\bf 3_1}$
is factorized:
\begin{eqnarray}
&&\hat{m}^2(1-q\hat{m}^4)(1-q^3\hat{m}^4)
P_{\bf 3_1}(\hat{l},\hat{m})
\nonumber \\
&&
=(1-q\hat{m}^4)\hat{l}^2
-q^{1/2}(1-q^3\hat{m}^4-q^4\hat{m}^6+q^5\hat{m}^{10})
-q^2(1-q^3\hat{m}^4)\hat{m}^6
\nonumber \\
&&=\hat{l}(1-q^{-1}\hat{m}^4)\hat{l}-q^{1/2}(1-q^3\hat{m}^4)\hat{l}
+q^{3/2}\hat{l}(1-q^{-1}\hat{m}^4)\hat{m}^6-q^2(1-q^3\hat{m}^4)\hat{m}^6
\nonumber \\
&&=\bigl[
\hat{l}(1-q^{-1}\hat{m}^4)-q^{1/2}(1-q^3\hat{m}^4)
\bigr]
(\hat{l}+q^{3/2}\hat{m}^6).
\end{eqnarray}

There are two branches for the solution of this difference equation.
One branch corresponds to the solution $l=-m^6$ for $P_{\bf
  3_1}(l,m^2)=0$ in $q\to 1$ limit, and we call this the {\it
  non-abelian branch}.  Another branch corresponds to a solution
$l=1$, and we call this the {\it abelian branch}.  In the following,
we will discuss the non-abelian branch for the trefoil knot.  The
results of abelian branch and the other torus knots are summarized in
appendix A.

The perturbative invariants $S_n^{(\alpha)}(u)$ for the branch $\alpha$
are defined as follows:
\begin{eqnarray}
W_{n}(K;q)=\exp\left[
\frac{1}{\hbar}S_0^{(\alpha)}(u)-\frac{1}{2}\delta^{(\alpha)}\log\hbar
+\sum_{n=0}^{\infty}S_{n+1}^{(\alpha)}(u)\hbar^n
\right].
\label{Wilson}
\end{eqnarray}
In the non-abelian branch $\alpha={\rm nab}$, the leading term yields \cite{KK,Witten}
\begin{eqnarray}
S_0^{\rm (nab)}(u)&=&{\rm CS}({\bf 3_1};0)-\int_{0}^u du\;v_{\rm nab}(u)
+\pi i u \nonumber  \\
&=&3\log^2 m-\frac{\pi^2}{12}-2\pi^2 s,
\quad s\in \mathbb{Z},
\\
l&=&-m^6.
\end{eqnarray}
The perturbative invariant for this branch satisfies the 
$q$-difference equation:
\begin{eqnarray}
(\hat{l}+q^{3/2}\hat{m}^6)
W^{\rm (nab)}_{n}({\bf 3_1};q)=0.
\end{eqnarray}
From this $q$-difference equation, we obtain the perturbative
invariants in the non-abelian branch:
\begin{eqnarray}
&&
S_0^{\rm (nab)\;\prime}(u)=6\log m, \\
&&
S_n^{\rm (nab)}(u)={\rm constant}, \quad n\ge 1.
\end{eqnarray}
Since the non-abelian branch corresponds to the geometric branch 
for the hyperbolic knots, we are able to compare our result with the BKMP's free energy
computed from the Eynard-Orantin topological recursion.

\subsection{Free energies on the character variety for the $(p,q)$ torus knot}
The character variety ${\cal C}^{(p,q)}$ corresponding to the non-abelian branch of the $(p,q)$ torus knot is given by
\begin{equation}
{\cal C}^{(p,q)}=\left\{x,y \in {\IC}^*~|~A(x,y)=y+x^n=0,~n:=pq \right\}.
\label{eqn:1}
\end{equation}
By making use of the topological recursion relation (\ref{eqn:3.2}), let us compute the free energies ${\cal F}^{(g,h)}(x_1,\ldots,x_h)$ defined in (\ref{eqn:3.46}) on the character variety ${\cal C}^{(p,q)}$. In this appendix, for simplicity at first we do not introduce the averaged meromorphic differentials ${\widetilde W^{(g,h)}}(x_1,\ldots,x_h)$, instead we use $W^{(g,h)}(x_1,\ldots,x_h)$:
\begin{eqnarray}
\label{eqn:2}
&&{\cal F}^{(g,h)}(x_1,\ldots,x_h):=\int^{x_1} \cdots \int^{x_h}{\widehat W}^{(g,h)}(x_1',\ldots,x_h'),\hspace{13em}\\
\label{eqn:3}
&&{\widehat W}^{(0,1)}(x):=\log y(x)\frac{dx}{x},\quad {\widehat W}^{(0,2)}(x_1,x_2):=B(x_1,x_2)-\frac{dy_1dy_2}{(y_1-y_2)^2},~y_i:=y(x_i),\\
\label{eqn:4}
&&{\widehat W}^{(g,h)}(x_1,\ldots,x_h):=W^{(g,h)}(x_1,\ldots,x_h)~ \mbox{for}~(g,h)\neq (0,1),(0,2),
\end{eqnarray}
and after the computation we take the average. Here we treat $y(p)=y({\bar p})$ as the projected coordinate on ${\cal C}^{(p,q)}$. Since (\ref{eqn:1}) has no ramification point, we introduce a free parameter $\mu$ as\footnote{When $\mu=-1$, and $n=1$, the curve ${\widetilde A}(x,y)=0$ is nothing but the Lambert curve, and then ${\widehat W^{(g,h)}}(x_1,\ldots,x_h)$ gives a generating function of the Hurwitz numbers \cite{BM,EMSf}.}:
\begin{equation}
{\widetilde {\cal C}}^{(p,q)}=\left\{x,y \in {\IC}^*~|~{\widetilde A}(x,y)=y+x^ne^{n\mu x}=0,~n=pq \right\},
\label{eqn:9}
\end{equation}
and from $\partial_x {\widetilde A}(x,y)=0$, we find that the deformed curve ${\widetilde {\cal C}}^{(p,q)}$ has one ramification point:
\begin{equation}
(x,y)=(-\mu^{-1},-(-\mu)^{-n}e^{-n}).
\label{eqn:10}
\end{equation}
On the curve ${\widetilde {\cal C}}^{(p,q)}$, the Bergman kernel $B(x_1,x_2)$ and the third type differential $dE_{p,\bar{p}}(x_1)$ are given by
\begin{eqnarray}
\label{eqn:11}
&&B(x_1,x_2)=\frac{dx_1dx_2}{(x_1-x_2)^2},\\
\label{eqn:12}
&&-dE_{p,\bar{p}}(x_1)=\frac12 \int_{{\bar p}}^{p} B(x_1,\xi)=\frac{dx_1}{2}\bigg\{\frac{1}{x_1-x(p)}-\frac{1}{x_1-x({\bar p})}\bigg\}.
\end{eqnarray}
To solve the recursion (\ref{eqn:3.2}) we have to consider the expansion of $W^{(g,h)}(x_1,\ldots,x_h)$ around the ramification point (\ref{eqn:10}), and for the purpose we introduce a parameter $p=\zeta$ near the ramification point as \cite{BKMP,BM}:
\begin{equation}
x(\zeta)=-\mu^{-1}+\zeta,\quad x({\bar \zeta})=-\mu^{-1}+S(\zeta),\quad S(\zeta):=-\zeta+\sum_{k=2}^{\infty}C_k\zeta^k,
\label{eqn:13}
\end{equation}
where $C_k$ are iteratively determined by the equation:
\begin{equation}
y(\zeta)=-x(\zeta)^ne^{n\mu x(\zeta)}=-x({\bar \zeta})^ne^{n\mu x({\bar \zeta})}=y({\bar \zeta}).
\label{eqn:14}
\end{equation}
From this equation we obtain the algebraic equation for $C_k$:
\begin{equation}
(1-\mu \zeta)e^{\mu \zeta}=(1-\mu S(\zeta))e^{\mu S(\zeta)}.
\label{eqn:15}
\end{equation}
Here we rescale the parameters $x$ and $\zeta$ as ${\widetilde x}:=\mu x$ and ${\widetilde \zeta}:=\mu \zeta$ respectively, and then from the equation (\ref{eqn:15}) we find
\begin{equation}
-\zeta S(\zeta):=-{\widetilde S}({\widetilde \zeta})={\widetilde \zeta}+\frac{2}{3}{\widetilde \zeta}^2+\frac{4}{9}{\widetilde \zeta}^3+\frac{44}{135}{\widetilde \zeta}^4+\frac{104}{405}{\widetilde \zeta}^5+\frac{40}{189}{\widetilde \zeta}^6+\frac{7648}{42525}{\widetilde \zeta}^7+\cdots.
\label{eqn:16}
\end{equation}

The annulus amplitude ${\cal F}^{(0,2)}(x_1,x_2)$ yields
\begin{eqnarray}
\label{eqn:17}
&&
{\cal F}^{(0,2)}(x_1,x_2)=\int \int \frac{dx_1dx_2}{(x_1-x_2)^2}-\frac{dy(x_1)dy(x_2)}{(y(x_1)-y(x_2))^2}~
\mathop{\longrightarrow}^{\mu \to 0}~\log\frac{-1}{\sum_{k=0}^{n-1}x_1^{n-1-k}x_2^k}, \hspace{3em}\\
&&
{\cal F}^{(0,2)}(x,x)~\mathop{\longrightarrow}^{\mu \to 0}~(1-n)\log x+\mbox{const.}
\label{eqn:18}
\end{eqnarray}
If we consider the averaged Bergman kernel as in (\ref{eqn:3.48}), then the averaged annulus amplitude has the form ${\cal F}^{(0,2)}(x)=\log\frac{x-x^{-1}}{x^n-x^{-n}}$.

Next we compute the higher free energies by the recursions. Using
\begin{equation}
y(\zeta)\Big(\frac{dy(\zeta)}{d\zeta}\Big)^{-1}=y(\zeta)\Big(\frac{dy(x)}{dx}\frac{dx(\zeta)}{d\zeta}\Big)^{-1}=\frac{\zeta-\mu^{-1}}{n\mu \zeta},
\label{eqn:19}
\end{equation}
and (\ref{eqn:15}), we find
\begin{equation}
\frac{dE_{\zeta,\bar{\zeta}}(x_1)}{\omega(\zeta)-\omega({\bar \zeta})}=\frac{({\widetilde \zeta}-1)d{\widetilde x}_1}{2n{\widetilde \zeta}({\widetilde \zeta}-{\widetilde S}({\widetilde \zeta})) d{\widetilde \zeta}}\Big\{\frac{1}{{\widetilde x}_1+1-{\widetilde \zeta}}-\frac{1}{{\widetilde x}_1+1-{\widetilde S}({\widetilde \zeta})}\Big\}.
\label{eqn:20}
\end{equation}
Using (\ref{eqn:3.11}) and (\ref{eqn:20}), we can compute ${\cal F}^{(0,3)}(x_1,x_2,x_3)$ as:
\begin{eqnarray}
{\cal F}^{(0,3)}(x_1,x_2,x_3)&& \nonumber\\
&&\hspace{-6em}
=\int \mathop{\mbox{Res}}_{{\widetilde \zeta}=0} \frac{({\widetilde \zeta}-1){\widetilde S}'({\widetilde \zeta})d{\widetilde \zeta} d{\widetilde x}_1d{\widetilde x}_2d{\widetilde x}_3}{2n {\widetilde \zeta}({\widetilde \zeta}-{\widetilde S}({\widetilde \zeta}))({\widetilde x}_2+1-{\widetilde \zeta})^2({\widetilde x}_3+1-{\widetilde S}({\widetilde \zeta}))^2}\Big\{\frac{1}{{\widetilde x}_1+1-{\widetilde \zeta}}-\frac{1}{{\widetilde x}_1+1-{\widetilde S}({\widetilde \zeta})}\Big\} \nonumber\\
&&\hspace{-6em}
=
\int\frac{1}{n}\prod_{i=1}^3 \frac{d{\widetilde x}_i}{({\widetilde x}_i+1)^2}=-\frac{1}{n}\prod_{i=1}^3 \frac{1}{(\mu x_i+1)^2}\quad \mathop{\longrightarrow}^{\mu \to 0}\quad -\frac{1}{n},
\label{eqn:21}
\end{eqnarray}
and thus ${\cal F}^{(0,3)}(x_1,x_2,x_3)$ is constant on ${\cal C}^{(p,q)}$. Using (\ref{eqn:3.12}) and (\ref{eqn:20}), we can also compute ${\cal F}^{(1,1)}(x_1)$ as:
\begin{eqnarray}
{\cal F}^{(1,1)}(x_1)&=&\int \mathop{\mbox{Res}}_{{\widetilde \zeta}=0}
\frac{({\widetilde \zeta}-1){\widetilde S}'({\widetilde \zeta})}{2n{\widetilde \zeta}({\widetilde \zeta}-{\widetilde S}({\widetilde \zeta}))^3}\Big\{\frac{1}{{\widetilde x}_1+1-{\widetilde \zeta}}-\frac{1}{{\widetilde x}_1+1-{\widetilde S}({\widetilde \zeta})}\Big\}d{\widetilde \zeta} d{\widetilde x}_1 \nonumber\\
&=&
-\frac{1}{24n}\int\frac{{\widetilde x}_1({\widetilde x}_1+4)}{({\widetilde x}_1+1)^4}d{\widetilde x}_1=\frac{1}{24n}\frac{\mu^2 x_1^2+3\mu x_1+1}{(\mu x_1+1)^3}\quad \mathop{\longrightarrow}^{\mu \to 0}\quad \frac{1}{24n}.
\label{eqn:22}
\end{eqnarray}
We see that the free energy ${\cal F}^{(1,1)}(x_1)$ is also constant on
${\cal C}^{(p,q)}$. In the same way, the free energies ${\cal
F}^{(0,4)}(x_1,x_2,x_3,x_4)$ and ${\cal F}^{(1,2)}(x_1,x_2)$ on
${\widetilde {\cal C}}^{(p,q)}$ are computed by (\ref{eqn:3.13}) and
(\ref{eqn:3.14}), and we find ${\cal F}^{(0,4)}(x_1,x_2,x_3,x_4)={\cal
F}^{(1,2)}(x_1,x_2)=0$ on ${\cal C}^{(p,q)}$. 
One can easily find that ${\widehat W}^{(g,h)}(x_1,\ldots,x_h)$ are
expressed by the rescaled variables ${\widetilde x}_i$, and
therefore we see that ${\cal F}^{(g,h)}(x_1,\ldots,x_h)$ are also constant on
${\cal C}^{(p,q)}$ after taking the average. This matches the
result that the asymptotic expansion of the Wilson loop expectation
value $W_n(K;q)$ along the $(p,q)$ torus knot is trivial on the non-abelian
branch.\footnote{
For annulus free energy ${\cal F}^{(0,2)}(x)$, we find the non-trivial
contribution $-\log (x^{n-1}+x^{n-2}+\cdots+x^{-n+2}+x^{-n+1})$ 
even after taking $\mu\to 0$ limit. 
In the WKB expansion of $W_n(T_{p,q};q)$, the perturbative invariant $S_1(u)$
vanishes in the non-abelian branch.
Here we consider this discrepancy would come from the normalization
factor of the partition function. 
}
The constants of the higher order
terms $S^{({\rm nab})}_n$ $(n\ge 2)$ may also come from the end points of the 
integration of the BKMP's free energies, although we do not know the
correct prescription to determine them rigorously at present. In this
computation, we found the triviality of the $u$-dependence of the  
perturbative invariants and BKMP's free energy for the torus knot.

\section{Conclusions and Discussions}

In this paper, we have discussed the correspondence between the
perturbative invariants of $SL(2;\mathbb{C})$ Chern-Simons gauge
theory and the free energies of the topological string defined
{\it\`a la} BKMP on the character variety for the figure eight knot
complement, the once punctured torus bundle over $\mathbb{S}^1$ with
the holonomy $L^2R$, and the $(p,q)$ torus knots.  On the three dimensional
geometry side, we computed the perturbative expansion of the
partition function of the state integral model around the saddle
point which corresponds to the geometric branch for the figure eight
knot complement, the once punctured torus bundle over $\mathbb{S}^1$
with the holonomy $L^2R$.  For the torus knots, we adopted the
factorized $q$-difference equation for the colored Jones polynomial.
On the character variety side, we computed the free energies on the
basis of the Eynard-Orantin topological recursion. We found the
coincidence to the fourth order on both sides under some particular
regularization of $G^n$ in the Bergman kernel.

The most ambiguous point in our discussion is the regularization of
the constants $G^n$ for each $n$ independently, although we found a
nice presentation for the regularization. Without this regularization,
we cannot establish an exact coincidence. But in the free energy
computations, there exists an ambiguity of the choice of the
integration path. In this paper we have picked-up the end points of
the integrations and neglected the contribution from the reference
points $(u_*,v_*)$.  In the context of the volume conjecture, the
analytic continuation is discussed in detail in the recent work
\cite{Witten}.  The Stokes phenomenon is also applicable to determine
the higher order terms in the WKB expansion, so further study along
these lines may fix the ambiguity completely.

In \cite{DGH}, the relation between the Chern-Simons gauge theory 
on 3-manifold $M$ and the two dimensional ${\cal N}=(2,2)$ theory on
$\mathbb{R}_{\hbar}$ is discussed via five dimensional ${\cal N}=2$
supersymmetric gauge theory. The analogous relation is
discussed in the AGT correspondence which connects four dimensional 
${\cal N}=2$ $SU(2)$ supersymmetric gauge theory 
and two dimensional Liouville field theory \cite{Alday:2009aq}.
In the context of the AGT correspondence, the surface operator in the four
dimensional ${\cal N}=2$ gauge theory can be realized by the non-compact
toric brane in the geometric engineering \cite{KPW,DGH,AFKMY}.
There may exist some relations between the chiral boson theory
\cite{DVKS} on the character variety and the two dimensional ${\cal N}=(2,2)$ 
 \cite{DFM2}.

\vspace{2em}

\noindent{\bf Acknowledgements:}

The authors would like to thank Andrea Brini, Sergio Cecotti, Tudor
Dimofte, Sergei
Gukov, Marcos Mari\~no, Masanori Morishita, Hitoshi Murakami, Yuji
Terashima, and Cumrun Vafa for fruitful discussions and useful
comments.  R.D. wishes to thank the Simons Center for Geometry and
Physics for providing a stimulating environment and generous
hospitality, and the participants of the 2010 Simons Workshop in
Mathematics and Physics for interesting discussions.  Two of the
authors (H.F. and R.D.) are also grateful to RIKEN and IPMU for warm
hospitality. The work of H.F. and M.M. is supported by the
Grant-in-Aid for Nagoya University Global COE Program, {\it Quest for
  Fundamental Principles in the Universe: from Particles to the Solar
  System and the Cosmos}.
H.F. is also supported by Grant-in-Aid for Young Scientists (B)
[\#21740179] from the Japan Ministry of Education, Culture, Sports,
Science and Technology.  The research of R.D. is supported by a NWO
Spinoza grant and the FOM program {\it String Theory and Quantum
  Gravity}.


\appendix
\section{Perturbative invariants from AJ conjecture}
In this appendix, we will summarize some computations on AJ conjecture.

\subsection{Factorization of AJ conjecture}
The AJ conjecture is the $q$-difference equation for the colored Jones 
polynomial. The factorization of the $q$-difference equation will occur for any knots \cite{Dimofte}.
In the following, we will see such factorization explicitly for the 
figure eight knot.

For the figure eight knot, the $q$-difference equation
yields \cite{Garou-Le, Geronimo}:
\begin{eqnarray}
&&A_q(\hat{l},\hat{m})J_n(K;q)=0,\quad
A_q(\hat{l},\hat{m})=\sum_{j=0}^3a_j(\hat{m};q)l^{j},\\
&&a_0(\hat{m};q)=\frac{q^5\hat{m}^2(-q^3+q^3\hat{m}^2)}{(q^2+q^3\hat{m}^2)(-q^5+q^6\hat{m}^4)},\nonumber \\
&&
a_1(\hat{m};q)=-\frac{q^2-q^3\hat{m}^2
}{q^5\hat{m}^2(q+q^3\hat{m}^2)(q^5-q^6\hat{m}^4)}
\nonumber \\
&&\hspace*{2cm}\times
(q^8-2q^9\hat{m}^2+q^{10}\hat{m}^2-q^{9}\hat{m}^4+q^{10}\hat{m}^4-q^{11}\hat{m}^4+q^{10}\hat{m}^6-2q^{11}\hat{m}^6+q^{12}\hat{m}^8),
\nonumber \\
&&a_2(\hat{m};q)=\frac{-q+q^3\hat{m}^2}{q^4\hat{m}^2(q^2+q^3\hat{m}^2)(-q+q^6\hat{m}^4)}
\nonumber\\
&&\hspace*{2cm}\times
(q^4+q^5\hat{m}^2-2q^{6}\hat{m}^2-q^{7}\hat{m}^4+q^{8}\hat{m}^4-q^{9}\hat{m}^4-2q^{10}\hat{m}^6+q^{11}\hat{m}^6+q^{12}\hat{m}^8),
\nonumber \\
&&a_3(\hat{m};q)=\frac{q^4\hat{m}^2(-1+q^3\hat{m}^2)}{(q+q^3\hat{m}^2)(q-q^6\hat{m}^4)},\nonumber 
\end{eqnarray}
where $J_n(K;q)$ is the colored Jones polynomial.
The $q$-Weyl operators $(\hat{m},\hat{l})$ satisfies
\begin{eqnarray}
&&\hat{m}f(u)=e^{u} f(u), \quad \hat{l}f(u)=f(u+\hbar), \\
&&\hat{l}\hat{m}=q^{1/2}\hat{m}\hat{l},\quad q=e^{2\hbar}.
\end{eqnarray}
The Jones polynomial is normalized as $J({\rm unknot};q)=1$.
Taking into account for the normalizations of the
colored Jones polynomial and the $SL(2,\mathbb{C})$ Chern-Simons partition function, 
one finds the $q$-difference equation for the $SL(2,\mathbb{C})$ Chern-Simons
partition function $Z_{\hbar}(M,u;q)$ as follows \cite{DGLZ}:
\begin{eqnarray}
&&\tilde{A}_q(\hat{l},\hat{m})Z_{\hbar}(M,u;q)=0,\quad 
\tilde{A}_q(\hat{l},\hat{m})=\sum_{j=0}^3\tilde{a}_j(\hat{m};q)l^{j},
\\
&&\tilde{a}_0(\hat{m};q)=\frac{q\hat{m}^2}{(1+q\hat{m}^2)(-1+q\hat{m}^4)},
\nonumber \\
&&\tilde{a}_1(\hat{m};q)=\frac{1+(q^2-2q)\hat{m}^2-(q^3-q^2-q)\hat{m}^4-(2q^3-q^2)\hat{m}^6+q^4\hat{m}^8}{q^{1/2}\hat{m}^2(1+q^2\hat{m}^2-q\hat{m}^4-q^3\hat{m}^6)},\nonumber \\
&& \tilde{a}_2(\hat{m};q)=-\frac{1-(2q^2-q)\hat{m}^2-(q^5-q^4-q^3)\hat{m}^4+(q^7-2q^6)\hat{m}^6+q^8\hat{m}^8}{q\hat{m}^2(1+q\hat{m}^2-q^5\hat{m}^4-q^6\hat{m}^6)},\nonumber \\
&& \tilde{a}_0(\hat{m};q)=-\frac{q^4\hat{m}^2}{q^{1/2}(1+q\hat{m}^2)(-1+q^5\hat{m}^4)}.\nonumber 
\end{eqnarray}
In $q\to 1$ limit, $\hat{A}_q(\hat{l},\hat{m})$ yields the
A-polynomial. But the abelian part $(l-1)$ is included.
We can show that this abelian part is factorizable even for the 
$q$-difference operator as:\footnote{
The partition function of the state integral model satisfies
the factored $q$-difference equation \cite{Dimofte}.
}
\begin{eqnarray}
&& \tilde{A}_q(\hat{l},\hat{m})=(q^{1/2}\hat{l}-1)\hat{A}_q(\hat{l},\hat{m}),\\
&&
 \hat{A}_q(\hat{l},\hat{m})
=\frac{q\hat{m}^2}{(1+q\hat{m}^2)(-1+q\hat{m}^4)}
-\frac{(-1+q\hat{m}^2)(1-q\hat{m}^2-(q+q^3)\hat{m}^4-q^3\hat{m}^6+q^4\hat{m}^8)}{q^{1/2}\hat{m}^2(-1+q\hat{m}^4)(-1+q^3\hat{m}^4)}\hat{l}
\nonumber \\
&&\hspace{2cm}
+\frac{q^{2}\hat{m}^2}{(1+q\hat{m}^2)(-1+q^3\hat{m}^4)}\hat{l}^2.
\end{eqnarray}
From this factorization, we expect that the AJ conjecture will imply the 
quantum Riemann surface structure in topological string theory \cite{DHS1,DHS2}.

\subsection{Abelian branch}
We will discuss the perturbative invariants near the abelian branch from
AJ conjecture. For the figure eight knot, the abelian branch is studied
\cite{DGLZ}.
In particular for the torus knots, the abelian branch contains rich structure
rather than the non-abelian branch.
One of the outstanding properties of this branch will be
the Melvin-Morton-Rozansky conjecture \cite{MMR,Mor,Roz1,GL}.
Here we discuss the expansion of $W_n(K;q)$ near the abelian branch point.

The leading term of the perturbative invariant (\ref{Wilson}) 
for the trefoil knot ${\bf 3_1}$ in this branch yields 
\begin{eqnarray}
&&l=1,\\
&&S_0^{({\rm abel})\;\prime}(u)=0.
\end{eqnarray}
Adopting this initial condition into the $q$-difference equation, one
finds a non-trivial expansion:
\begin{eqnarray}
&&\hspace*{-1cm}
S_1^{\rm (abel)}(u)=\log\frac{m(m^2-1)}{m^4-m^2+1}, 
\label{Reidemeister}
\\
&&\hspace*{-1cm}
S_2^{\rm (abel)}(u)=\frac{2m^4}{(1-m^2+m^4)^2},\\
&&\hspace*{-1cm}
S_3^{\rm (abel)}(u)=-\frac{2m^4(1-4m^4+m^8)}{(1-m^2+m^4)^4}, \\
&&\hspace*{-1cm}
S_4^{\rm (abel)}(u)=\frac{4m^4(1+2m^2-23m^4-4m^6+60m^8-4m^{10}-23m^{12}+2m^{14}+m^{16})}{3(1-m^2+m^4)^6}.
\end{eqnarray}
The partition function in this branch has the polynomial growth,
since $S_0^{({\rm abel})}(u)=0$. 
The volume conjecture for the torus knots 
in this branch is studied in \cite{KT,DK,HM2}.
The perturbative solution above is consistent with \cite{HM2,Roz2}.
In particular, the subleading term $S_1^{({\rm abel})}(u)$ is 
\begin{eqnarray}
e^{S_1^{(\rm abel)}}=\frac{2\sinh(u/2)}{\Delta(T_{p,q};m)},
\label{KTexp}
\end{eqnarray}
where $\Delta(T_{p,q};m)$ is Alexander polynomial. 
In the case of trefoil knot, the Alexander polynomial is 
$\Delta(K;m)=m^2+m^{-2}-1$, and
this result is consistent with (\ref{Reidemeister}).

\subsection{The other examples of torus knots}
From (\ref{inhom_2_2m+1}), one can also find the perturbative invariants
for the torus knots in each branch. Here we will show some computational results for
$(2,5)$ and $(2,7)$ torus knots.

\noindent\underline{\it $\bullet$ (2,5) torus knot}\\
The $q$-difference equation for the cinquefoil knot is
\begin{eqnarray}
&&\sum_{j=0}^2a_j(q^{n/2};q)W_{n+j}(T_{2,5};q)=0,
\label{AJ_5_2}
\\
&&
a_0(\hat{m},q)=\hat{m}^{10} q^{11} (-1 + \hat{m}^4 q^3),
\nonumber \\
&&
a_1(\hat{m},q)
=q^{17/2} (-1 + \hat{m}^4 q^3 + \hat{m}^{10} q^6 -\hat{m}^{14} q^7),
\nonumber \\
&& 
a_2(\hat{m},q)
=q^7 - \hat{m}^4 q^8.
\nonumber 
\end{eqnarray}
The $q$-difference operator 
\begin{eqnarray}
\hat{A}_{T_{2,2\mathfrak{m}+1}}(\hat{l},\hat{m})=\sum_{j=0}^2a_j(\hat{m},q)\hat{l}^j, 
\end{eqnarray}
is factorized for $\mathfrak{m}=2$ as follows:
\begin{eqnarray}
\hat{A}_{T_{2,5}}(\hat{l},\hat{m})=q^7[(\hat{l}(1-q^{-1}\hat{m}^4)-q^{3/2}(1-q^3\hat{m}^4)](\hat{l}+q^{5/2}\hat{m}^{10}).
\label{AJ_5_2_2}
\end{eqnarray}

In the abelian branch, one obtains the perturbative invariants 
for (2,5) torus knot as follows:
\begin{eqnarray}
&&l=1, \quad 
S_0^{({\rm abel})\;\prime}(u)=\log l, 
 \\
&&S_1^{({\rm abel})}(u)=\log\frac{m^3(m^2-1)}{1 - m^2 + m^4 - m^6 + m^8},
\label{Reidemeister_5_2}
\\
&&
S_2^{({\rm abel})}(u)=\frac{2 m^4 (1 - 2 m^2 + 4 m^4 - 2 m^6 + m^8)}{(1 - m^2 + m^4 -
m^6 + m^8)^2},
\\
&&
S_3^{({\rm abel})}(u)=-2 m^4 (1 - 4 m^2 + 12 m^4 - 12 m^6 - 6 m^8 + 32 m^{10} - 52 m^{12} + 
   32 m^{14} 
- 6 m^{16}
\nonumber \\
&&
\quad\quad\quad\quad\quad\quad\quad\quad
- 12 m^{18} + 12 m^{20} - 4 m^{22} + m^{24})
/(1 - m^2 + m^4 - m^6 + m^8)^4,
 \\
&&
S_4^{({\rm abel})}(u)=4 m^4 (1 - 6 m^2 + 25 m^4 - 16 m^6 - 165 m^8 + 602 m^{10} - 
    1141 m^{12} + 940 m^{14} 
\nonumber \\
&&\quad\quad\quad\quad\quad\quad\quad\quad
+ 449 m^{16}
 - 2330 m^{18} + 3342 m^{20} 
-   2330 m^{22} + 449 m^{24} + 940 m^{26} 
\nonumber \\
&&\quad\quad\quad\quad\quad\quad\quad\quad
- 1141 m^{28} 
+ 602 m^{30} 
- 165 m^{32} - 16 m^{34} + 25 m^{36} - 6 m^{38} + m^{40})
\nonumber \\
&&\quad\quad\quad\quad\quad\quad
/3 (1 - m^2 + m^4 - m^6 + m^8)^6.
\end{eqnarray}
The Alexander polynomial for $(2,5)$ torus knot is
\begin{eqnarray}
A(t)=1 - t + t^2 - t^3 + t^4,
\end{eqnarray}
and the $S_1^{({\rm abel})}(u)$ in (\ref{Reidemeister_5_2}) is consistent
with the general formula (\ref{KTexp}).

In the non-abelian branch, we find the trivial perturbative invariants
for the $(2,5)$ torus knot as follows:
\begin{eqnarray}
&&l=-m^{10}, \quad 
S_0^{({\rm nab})\prime}(u)=\log l, 
 \\
&& S_k^{({\rm nab})}(u)={\rm constant}, \quad {\rm for}\;\; k\ge 1.
\end{eqnarray}
The factorization of (\ref{AJ_5_2_2}) indicates the triviality of the
higher order terms.

\noindent\underline{\it $\bullet$ (2,7) torus knot}\\
For the $(2,7)$ torus knot, the $q$-difference equation is
\begin{eqnarray}
&&\sum_{j=0}^2a_j(q^{n/2};q)W_{n+j}(T_{(2,7)};q)=0,
\label{AJ_7_2}\\
&&
a_0(\hat{m},q)=\hat{m}^{14} q^{15} (-1 + \hat{m}^4 q^3), 
\nonumber \\
&&
a_1(\hat{m},q)=q^{23/2} (-1 + \hat{m}^4 q^3 + \hat{m}^{14} q^8 -
\hat{m}^{18} q^9),
\nonumber \\
&&
a_2(\hat{m},q)=-q^9 (-1 + \hat{m}^4 q).
\nonumber 
\end{eqnarray}
The $q$-difference operator 
$\hat{A}_{T_{2,7}}(\hat{l},\hat{m})$
is factorized as follows:
\begin{eqnarray}
\hat{A}_{T_{2,7}}(\hat{l},\hat{m})=q^9[\hat{l}(1-q^{-1}\hat{m}^4)-q^{5/2}(1-q^3\hat{m}^4)][\hat{l}+q^{7/2}\hat{m}^{14}].
\label{AJ_7_2_2}
\end{eqnarray}

In the abelian branch, we find the perturbative invariants
iteratively from (\ref{AJ_7_2}):
\begin{eqnarray}
&&l=1, \quad 
S_0^{({\rm abel})\;\prime}(u)=\log l, 
 \\
&&S_1^{({\rm abel})}(u)=\log\frac{m^5(-1+m^2)}
{1 - m^2 + m^4 - m^6 + m^8 - m^{10} + m^{12}}, \\
&&
S_2^{({\rm abel})}(u)=\frac{2 m^4 (1 - 2 m^2 + 4 m^4 - 6 m^6 + 9 m^8 - 6 m^{10} + 4 m^{12} - 2 m^{14} + 
   m^{16})}{(1 - m^2 + m^4 - m^6 + m^8 - m^{10} + m^{12})^2}, \\
&&
S_3^{({\rm abel})}(u)=-[2 m^4 (1 - 4 m^2 + 12 m^4 - 28 m^6 + 58 m^8 - 72 m^{10} + 44 m^{12} + 
   24 m^{14} - 125 m^{16} 
\nonumber \\
&&\quad\quad\quad\quad\quad
+ 224 m^{18} - 280 m^{20} 
+ 224 m^{22} - 125 m^{24} + 
   24 m^{26} + 44 m^{28} - 72 m^{30} + 58 m^{32} 
\nonumber \\
&&\quad\quad\quad\quad\quad
- 28 m^{34} + 12 m^{36} - 
   4 m^{38} + m^{40})]
\nonumber \\
&&\quad\quad\quad\quad
/(1 - m^2 + m^4 - m^6 + m^8 - m^{10} + m^{12})^4, \\
&&
S_4^{({\rm abel})}(u)=-4 m^4 (-1 + 6 m^2 - 25 m^4 + 80 m^6 - 219 m^8 + 314 m^{10} + 
     125 m^{12} - 1756 m^{14} 
\nonumber \\
&&\quad\quad\quad\quad\quad
+ 5186 m^{16} - 10288 m^{18} 
+ 15482 m^{20} - 
     16468 m^{22} + 9609 m^{24} + 5318 m^{26} 
\nonumber \\
&&\quad\quad\quad\quad\quad
- 24780 m^{28} + 41862 m^{30} - 
     49058 m^{32} + 41862 m^{34} 
- 24780 m^{36} + 5318 m^{38} 
\nonumber \\
&&\quad\quad\quad\quad\quad
+ 9609 m^{40} - 
     16468 m^{42} + 15482 m^{44} - 10288 m^{46} + 5186 m^{48} - 1756
     m^{50} 
\nonumber \\
&&\quad\quad\quad\quad\quad
+ 
     125 m^{52} + 314 m^{54} - 219 m^{56} + 80 m^{58} - 25 m^{60} + 6 m^{62} - 
     m^{64})
\nonumber \\
&&\hspace*{1.5cm}
/3(1 - m^2 + m^4 - m^6 + m^8 - m^{10} + m^{12})^6.
\end{eqnarray}
The Alexander polynomial for the $(2,7)$ torus knot is
\begin{eqnarray}
A(t)=1 - t + t^2 - t^3 + t^4 - t^5 + t^6,
\end{eqnarray}
and the perturbative invariant $S_1^{({\rm abel})}(u)$ is consistent
with (\ref{KTexp}).

In the non-abelian branch, we find the trivial perturbative invariants
for $(2,7)$ torus knot as follows:
\begin{eqnarray}
&&l=-m^{14}, \quad 
S_0^{({\rm nab})\;\prime}(u)=\log l, 
 \\
&& S_k^{(\rm nab)}(u)={\rm constant}, \quad {\rm for}\;\; k\ge 1.
\end{eqnarray}

\noindent\underline{\it $\bullet$ (2,p) torus knots (Conjecture)}\\
From the computations for $p=3,5,7$, we can guess the 
$q$-difference equation for $(2,p)$ torus knots:
\begin{eqnarray}
\hat{A}_{T_{p,2}}(\hat{l},\hat{m})=
[\hat{l}(1-q^{-1}\hat{m}^4)-q^{p/2-1}(1-q^3\hat{m}^4)][\hat{l}+q^{p/2}\hat{m}^{2p}]=0.
\end{eqnarray}

\section{Derivation of (\ref{eqn:3.35}) and (\ref{eqn:3.36})}\label{sectionB}

Here we consider the case that the character variety is genus one curve with two sheets written as (\ref{eqn:3.37}), and describe the derivation of (\ref{eqn:3.35}) and (\ref{eqn:3.36}) in detail. For computing (\ref{eqn:3.13}) and (\ref{eqn:3.14}) let us expand $W^{(0,3)}(p_1,p_2,q),~W^{(0,3)}(p,q,q)$ and $W^{(1,1)}(q)$ around $q=q_i$. At first, from (\ref{eqn:3.33}) and (\ref{eqn:3.34}) we consider the expansion of the kernel differentials $\chi_j^{(1)}(q)$ and $\chi_j^{(2)}(q)$ around $q=q_i$. By (\ref{eqn:3.32}) the kernel differentials are obtained:
\begin{eqnarray}
\label{eqn:B.1}
\chi_j^{(1)}(q)&=&\frac{ds}{M_j\sigma_j'}\frac{1}{\sqrt{\sigma(q;q_i)}}\Big(G+\frac{2f(q,q_j)}{(q-q_j)^2}\Big),\\
\label{eqn:B.2}
\chi_j^{(2)}(q)&=&\frac{ds}{3M_j\sigma_j'}\frac{1}{\sqrt{\sigma(q;q_i)}}\frac{4\sigma(q;q_j)-\sigma'(q)}{(q-q_j)^2}-\Big(\frac{M_j'}{M_j}+\frac{\sigma_j''}{3\sigma_j'}\Big)\chi_j^{(1)}(q),
\end{eqnarray}
where $\sigma(q;q_i):=\sigma(q)/(q-q_i)$. Using the expressions when $j \neq i$, we find the expansions
\begin{eqnarray}
\label{eqn:B.3}
&&\chi_j^{(1)}(q)\simeq \frac{ds}{M_j\sigma_j'}\frac{1}{\sqrt{\sigma_i'}}\Big\{G+\frac{2f(q_i,q_j)}{(q_i-q_j)^2}+{\cal O}(s^2)\Big\},\quad s^2:=q-q_i,\hspace{10em}\\
\label{eqn:B.5}
&&\chi_j^{(2)}(q)\simeq -\frac{ds}{M_j\sigma_j'}\frac{1}{\sqrt{\sigma_i'}}\Big\{\frac{\sigma_i'}{3(q_i-q_j)^2}+\Big(\frac{M_j'}{M_j}+\frac{\sigma_j''}{3\sigma_j'}\Big)\Big(G+\frac{2f(q_i,q_j)}{(q_i-q_j)^2}\Big)+{\cal O}(s^2)\Big\},
\end{eqnarray}
and when $j=i$, we find the expansions
\begin{eqnarray}
\label{eqn:B.6}
&&\chi_i^{(1)}(q)\simeq \frac{ds}{M_i\sqrt{\sigma_i'}}\Big\{\frac{1}{s^2}+\frac{1}{\sigma_i'}\Big(G-\frac{1}{12}\sigma_i''\Big)-\frac{s^2}{4\sigma_i'^2}\Big(\sigma_i''G-\frac{5}{24}\sigma_i''^2+\frac{1}{3}\sigma_i'\sigma_i'''\Big)+{\cal O}(s^4)\Big\},\hspace{2.3em}\\
&&\chi_i^{(2)}(q)\simeq \frac{ds}{M_i\sqrt{\sigma_i'}}\Big\{\frac{1}{s^4}-\frac{1}{s^2}\Big(\frac{M_i'}{M_i}+\frac{\sigma_i''}{4\sigma_i'}\Big)-\frac{G}{\sigma_i'}\Big(\frac{M_i'}{M_i}+\frac{\sigma_i''}{3\sigma_i'}\Big) \nonumber\\
\label{eqn:B.8}
&&\hspace{9.5em}
+\frac{1}{12\sigma_i'^2}\Big(\frac{11}{24}\sigma_i''^2+\sigma_i'\sigma_i''\frac{M_i'}{M_i}-\frac{1}{3}\sigma_i'\sigma_i'''\Big)+{\cal O}(s^2)\Big\}.
\end{eqnarray}
By the expansions, $W^{(0,3)}(p_1,p_2,q),W^{(0,3)}(p,q,q)$ and $W^{(1,1)}(q)$ can be expanded around $s^2=q-q_i=0$ as,
\begin{eqnarray}
&&W^{(0,3)}(p_1,p_2,q)\simeq
\frac{ds}{2M_i\sqrt{\sigma_i'}}\Big\{M_i^2\sigma_i'\Big(\frac{1}{s^2}+\frac{1}{\sigma_i'}\big(G-\frac{1}{12}\sigma_i''\big)\Big)\chi_i^{(1)}(p_1)\chi_i^{(1)}(p_2)\nonumber\\
\label{eqn:B.9}
&&\hspace{10em}
+\sum_{j \neq i}M_iM_j\Big(G+\frac{2f(q_i,q_j)}{(q_i-q_j)^2}\Big)\chi_j^{(1)}(p_1)\chi_j^{(1)}(p_2)+{\cal O}(s^2)\Big\},\\
&&W^{(0,3)}(p,q,q)\nonumber\\
&&\hspace{3em}\simeq
\frac{dsds}{2}\Big\{\Big(\frac{1}{s^4}+\frac{2}{s^2\sigma_i'}\big(G-\frac{1}{12}\sigma_i''\big)+\frac{1}{\sigma_i'^2}\big(G^2-\frac23\sigma_i''G +\frac19\sigma_i''^2-\frac16\sigma_i'\sigma_i'''\big)\Big)\chi_i^{(1)}(p) \nonumber\\
\label{eqn:B.10}
&&\hspace{15em}
+\sum_{j \neq i}\frac{1}{\sigma_i'\sigma_j'}\Big(G+\frac{2f(q_i,q_j)}{(q_i-q_j)^2}\Big)^2\chi_j^{(1)}(p)+{\cal O}(s^2)\Big\},\\
&&W^{(1,1)}(q)\simeq
\frac{ds}{16M_i\sqrt{\sigma_i'}}\Big\{\frac{1}{s^4}+\frac{1}{s^2\sigma_i'}\Big(4G-\frac{7}{12}\sigma_i''-\sigma_i'\frac{M_i'}{M_i}\Big)+\frac{1}{\sigma_i'^2}\Big(4G^2-\big(\sigma_i''+\sigma_i'\frac{M_i'}{M_i}\big)G \nonumber\\
&&\hspace{22em}
+\frac{1}{12}\big(\frac{19}{24}\sigma_i''^2+\sigma_i'\sigma_i''\frac{M_i'}{M_i}-\frac13\sigma_i'\sigma_i'''\big)\Big)
 \nonumber\\
&&\hspace{3.5em}
+\sum_{j\neq i}\frac{M_i}{M_j\sigma_j'^2}\Big(-\frac{\sigma_i'\sigma_j'}{3(q_i-q_j)^2}
+\big(4G-\frac23\sigma_j''-\sigma_j'\frac{M_j'}{M_j}\big)\big(G+\frac{2f(q_i,q_j)}{(q_i-q_j)^2}\big)\Big)+{\cal O}(s^2)\Big\},\nonumber\\
&&\label{eqn:B.11}
\end{eqnarray}
and from (\ref{eqn:3.13}) and (\ref{eqn:3.14}), we obtain (\ref{eqn:3.35}) and (\ref{eqn:3.36}).

\section{Computation of the subleading term ${\cal F}_1(p)$}

In this appendix, on the character variety
\begin{equation}
y(p)=M(p)\sqrt{\sigma(p)},\quad \sigma(p)=\prod_{i=1}^4(p-q_i)=p^4-S_1p^3+S_2p^2-S_1p+1,
\label{eqn:C.2}
\end{equation}
we compute (\ref{eqn:3.48}),
\begin{equation}
{\cal F}_1(p)=\frac12{\cal F}^{(0,2)}(p)=\frac14{\cal F}^{(0,2)}(p,p)=\frac12\int^{p} \int^{p}B(p_1',p_2')+B(p_1',p_2'^{-1})-\frac{dw_1'dw_2'}{(w_1'-w_2')^2},
\label{eqn:C.1}
\end{equation}
where $w_i'=(p_i'+p_i'^{-1})/2$. Using (\ref{eqn:3.28}) we get
\begin{eqnarray}
&&B(p_1,p_2)+B(p_1,p_2^{-1})\hspace{28em}\nonumber\\
\label{eqn:C.3}
&&\hspace{4em}=
\frac{dw_1dw_2}{2\sqrt{\widetilde{\sigma}(w_1)\widetilde{\sigma}(w_2)}(w_1-w_2)^2}\left(\sqrt{\widetilde{\sigma}(w_1)\widetilde{\sigma}(w_2)}+\widetilde{f}(w_1,w_2)\right)=:\widetilde{B}(w_1,w_2),\\
\label{eqn:C.4}
&&
\widetilde{\sigma}(w):=\frac{\sigma(p)}{p^2}=4w^2-2S_1w+(S_2-2)=4(w-\alpha_1)(w-\alpha_2),\\
\label{eqn:C.5}
&&
\widetilde{f}(w_1,w_2):=4w_1w_2-(w_1+w_2)S_1+(S_2-2),
\end{eqnarray}
where this is nothing but the Bergman kernel on the genus $0$ reduced curve $y^2=\widetilde{\sigma}(w)$. Here by a change of variable \cite{Eyn1},
\begin{equation}
2w(\lambda)=\frac{S_1}{2}+\gamma(\lambda+\lambda^{-1}),\quad \gamma:=\frac{\alpha_2-\alpha_1}{2},
\label{eqn:C.6}
\end{equation}
we can rewrite (\ref{eqn:C.4}) and (\ref{eqn:C.5}) as
\begin{eqnarray}
\label{eqn:C.7}
&&
\widetilde{\sigma}(w)=\gamma^2(\lambda-\lambda^{-1})^2,\\
&&
\widetilde{f}(w_1,w_2)=\gamma^2\Bigl((\lambda_1+\lambda_1^{-1})(\lambda_2+\lambda_2^{-1})-4\Bigr).
\label{eqn:C.8}
\end{eqnarray}
Therefore from (\ref{eqn:C.3}) we obtain
\begin{equation}
\widetilde{B}(w_1,w_2)-\frac{dw_1 dw_2}{(w_1-w_2)^2}=\frac{d\lambda_1 d\lambda_2}{(\lambda_1-\lambda_2)^2}-\frac{(\lambda_1^2-1)(\lambda_2^2-1)d\lambda_1 d\lambda_2}{(\lambda_1-\lambda_2)^2(\lambda_1\lambda_2-1)^2}=\frac{d\lambda_1 d\lambda_2}{(\lambda_1\lambda_2-1)^2},
\label{eqn:C.9}
\end{equation}
and then (\ref{eqn:C.1}) is easily computed as
\begin{eqnarray}
&&\frac14{\cal F}^{(0,2)}(p_1,p_2)=\frac12\int^{w_1}\int^{w_2} \widetilde{B}(w_1',w_2')-\frac{dw_1'dw_2'}{(w_1'-w_2')^2}\nonumber\\
&&\hspace{6.4em}=\frac12\int^{\lambda_1}\int^{\lambda_2}\frac{d\lambda_1' d\lambda_2'}{(\lambda_1'\lambda_2'-1)^2}=\frac12\log\frac{\lambda_2}{\lambda_1\lambda_2-1},\nonumber\\
&&
{\cal F}_1(p)=\frac{1}{2}\log\frac{1}{\lambda-\lambda^{-1}}=\frac{1}{2}\log \frac{\gamma}{\sqrt{{\widetilde \sigma}(w)}}.
\label{eqn:C.10}
\end{eqnarray}
This result coincides with the computation (\ref{S1Fig}) and (\ref{S1_m009}) after the identification of the parameter $w=(m^2+m^{-2})/2$ as discovered in \cite{DijFu}.

\section{Computation of the fifth order free energy ${\cal F}_4(p)$}\label{sectionF}
\newcommand{\eqnum}{\addtocounter{equation}{1}\tag*{(\normalsize{\ref{sectionF}.\arabic{equation}})}}

\begin{figure}[t]
 \begin{center}
  \includegraphics[width=160mm]{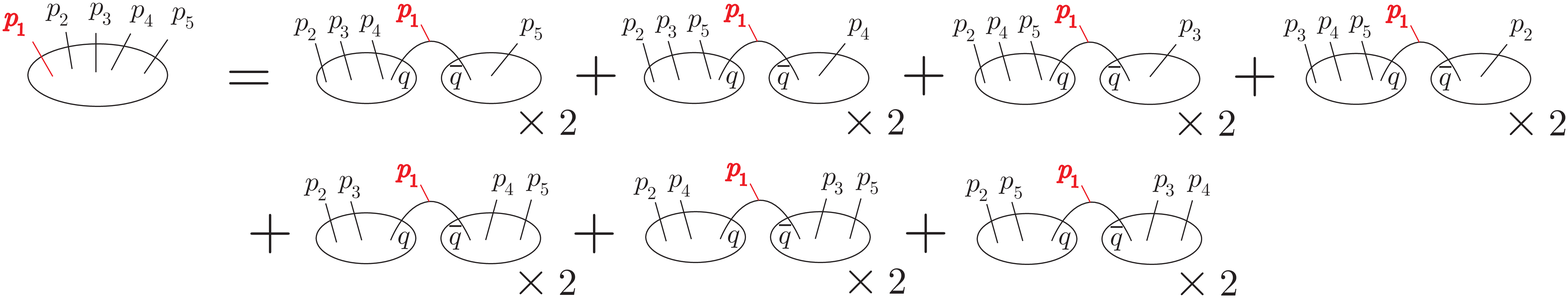}
 \end{center}
 \caption{Recursive structure of $W^{(0,5)}(p_1,\ldots,p_5)$}
 \label{fig:F1}
\end{figure}

By the recursion (\ref{eqn:3.2}), the multilinear meromorphic differentials with the Euler number $\chi=-3$ are obtained. Let us expand these differentials on the character variety (\ref{eqn:3.37}) in terms of the kernel differentials (\ref{eqn:3.15}). After some computation as in appendix {\ref{sectionB} we obtain the meromorphic differential $W^{(0,5)}(p_1,\ldots,p_5)$ as follows (see also Fig.\ref{fig:F1}):\vspace{-1em}

{\footnotesize
\begin{align*}
W^{(0,5)}(p_1,\ldots,p_5)&=\sum_{q_i\in{\cal C}}\mathop{\mbox{Res}}_{q=q_i}\frac{dE_{q,\bar{q}}(p_1)}{y(q)dq}\Big\{\big(B(q,p_2)W^{(0,4)}(\bar{q},p_3,p_4,p_5)+\mbox{perm}(p_2,p_3,p_4,p_5)\big) \\
&\hspace{9em}
+\big(W^{(0,3)}(q,p_2,p_3)W^{(0,3)}(\bar{q},p_4,p_5)+(p_3\leftrightarrow p_4)+(p_3\leftrightarrow p_5)\big)\Big\} \\
&
=\frac{3}{8}\sum_{i}\Big\{M_i^2\sigma'_i \big(5E_i^2+10(3A_i-a_i)E_i+24A_i^2+20B_i+\frac{10}{3}b_i \big)\chi_{i\hspace{0.15em}i\hspace{0.15em}i\hspace{0.15em}i\hspace{0.15em}i}^{11111}\\
&\hspace{1em}
+\sum_{j\neq i}M_iM_j \big(H_{ij}(E_j+3A_j-a_j)-\frac{\sigma_i'}{3(q_i-q_j)^2}\big)\big[\chi_{i\hspace{0.15em}i\hspace{0.05em}j\hspace{0.05em}j\hspace{0.05em}j}^{11111}+\mbox{perm}({}_5C_2)\big]\nonumber\\
&\hspace{1em}
+\sum_{j,k\neq i} \frac{M_jM_kH_{ij}H_{ik}}{3\sigma_i'}\big[\chi_{i\hspace{0.05em}j\hspace{0.05em}j\hspace{0.05em}k\hspace{0.05em}k}^{11111}+\mbox{perm}(\frac{5!}{2!2!2})\big]\\
&\hspace{1em}
+2M_i^2\sigma'_i\big(3E_i+10A_i\big)\big[\chi_{i\hspace{0.15em}i\hspace{0.15em}i\hspace{0.15em}i\hspace{0.15em}i}^{11112}+\mbox{perm}({}_5C_1)\big]
+\sum_{j\neq i}M_iM_jH_{ij}\big[\chi_{i\hspace{0.15em}i\hspace{0.05em}j\hspace{0.05em}j\hspace{0.05em}j}^{11112}+\mbox{perm}({}_4C_2\times 5)\big]\\
&\hspace{1em}
+6M_i^2\sigma'_i\big[\chi_{i\hspace{0.15em}i\hspace{0.15em}i\hspace{0.15em}i\hspace{0.15em}i}^{11122}+\mbox{perm}({}_5C_2)\big]
+5M_i^2\sigma'_i\big[\chi_{i\hspace{0.15em}i\hspace{0.15em}i\hspace{0.15em}i\hspace{0.15em}i}^{11113}+\mbox{perm}({}_5C_1)\big]\Big\}.\eqnum
\label{eqn:F4}
\end{align*}}
We have defined
\begin{eqnarray}
\label{eqn:F5}
&&
\chi_{\hspace{0.1em}i_1\hspace{0.2em}i_2\hspace{0.1em}\ldots \hspace{0.15em}i_h}^{n_1n_2\ldots n_h}:=\chi_{i_1}^{(n_1)}(p_1)\chi_{i_2}^{(n_2)}(p_2)\cdots\chi_{i_h}^{(n_h)}(p_h),\\
&&
a_i:=\frac{\sigma_i''}{3\sigma_i'}+\frac{M_i'}{M_i},\quad A_i:=\frac{\sigma_i''}{4\sigma_i'}+\frac{M_i'}{M_i},\quad b_i:=\frac{\sigma_i''^2}{32\sigma_i'^2}-\frac{\sigma_i'''}{12\sigma_i'},\nonumber\\
&&
B_i:=\frac{M_i''}{2M_i}+\frac{\sigma_i''M_i}{4\sigma_i'M_i}-b_i,\quad E_i:=\frac{1}{\sigma_i'}\big(G-\frac{1}{12}\sigma_i''\big),\quad H_{ij}:=G+\frac{2f(q_i,q_j)}{(q_i-q_j)^2},
\label{eqn:F6}
\end{eqnarray}
where $G=G(k)$, and $f(p,q)$ are defined in (\ref{eqn:3.29}), and (\ref{eqn:3.30}) respectively. In \ref{eqn:F4}, ``perm'' denotes the permutation of $p_1,\ldots, p_5$ so that the result becomes symmetric for these variables, for example,\vspace{-1em}

{\footnotesize
\begin{align*}
\chi_{i\hspace{0.05em}j\hspace{0.05em}j\hspace{0.05em}k\hspace{0.05em}k}^{11111}+\mbox{perm}(\frac{5!}{2!2!2})&=
\chi_{i\hspace{0.05em}j\hspace{0.05em}j\hspace{0.05em}k\hspace{0.05em}k}^{11111}+\chi_{j\hspace{0.05em}i\hspace{0.05em}j\hspace{0.05em}k\hspace{0.05em}k}^{11111}+\chi_{j\hspace{0.05em}j\hspace{0.05em}i\hspace{0.05em}k\hspace{0.05em}k}^{11111}+\chi_{k\hspace{0.05em}j\hspace{0.05em}j\hspace{0.05em}i\hspace{0.05em}k}^{11111}+\chi_{k\hspace{0.05em}j\hspace{0.05em}j\hspace{0.05em}k\hspace{0.05em}i}^{11111}+\chi_{i\hspace{0.05em}k\hspace{0.05em}j\hspace{0.05em}j\hspace{0.05em}k}^{11111}+\chi_{k\hspace{0.05em}i\hspace{0.05em}j\hspace{0.05em}j\hspace{0.05em}k}^{11111}\\
&
+\chi_{k\hspace{0.05em}j\hspace{0.05em}i\hspace{0.05em}j\hspace{0.05em}k}^{11111}+\chi_{k\hspace{0.05em}k\hspace{0.05em}j\hspace{0.05em}i\hspace{0.05em}j}^{11111}+\chi_{k\hspace{0.05em}k\hspace{0.05em}j\hspace{0.05em}j\hspace{0.05em}i}^{11111}+\chi_{i\hspace{0.05em}j\hspace{0.05em}k\hspace{0.05em}j\hspace{0.05em}k}^{11111}+\chi_{k\hspace{0.05em}i\hspace{0.05em}j\hspace{0.05em}k\hspace{0.05em}j}^{11111}+\chi_{j\hspace{0.05em}k\hspace{0.05em}i\hspace{0.05em}j\hspace{0.05em}k}^{11111}+\chi_{j\hspace{0.05em}k\hspace{0.05em}j\hspace{0.05em}i\hspace{0.05em}k}^{11111}+\chi_{j\hspace{0.05em}k\hspace{0.05em}j\hspace{0.05em}k\hspace{0.05em}i}^{11111}.
\eqnum
\label{eqn:F7}
\end{align*}}

\begin{figure}[t]
 \begin{center}
  \includegraphics[width=110mm]{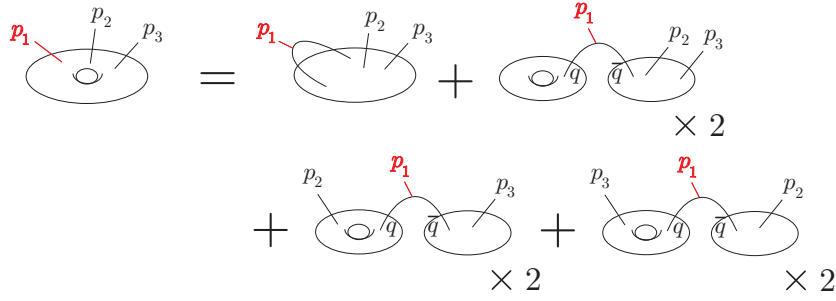}
 \end{center}
 \caption{Recursive structure of $W^{(1,3)}(p_1,p_2,p_3)$}
 \label{fig:F2}
\end{figure}
\noindent In the same way we obtain the meromorphic differential $W^{(1,3)}(p_1,p_2,p_3)$ as follows (see also Fig.\ref{fig:F2}):\vspace{-1em}

{\footnotesize
\begin{align*}
&
W^{(1,3)}(p_1,p_2,p_3)=\sum_{q_i\in{\cal C}}\mathop{\mbox{Res}}_{q=q_i}\frac{dE_{q,\bar{q}}(p_1)}{2y(q)dq}\Big\{W^{(0,4)}(q,\bar{q},p_2,p_3)+2W^{(1,1)}(q)W^{(0,3)}(\bar{q},p_2,p_3)\\
&\hspace{21em}
+2\big(B(q,p_2)W^{(1,2)}(\bar{q},p_3)+(p_2\leftrightarrow p_3)\big)\Big\} \\
&
=\frac{1}{64}\sum_{i}\Big\{\big(32E_i^3+(252A_i-195a_i)E_i^2+(522A_i^2-846A_ia_i+318a_i^2+120B_i+92b_i)E_i \\
&\quad
-30A_iB_i+332A_ib_i-262a_ib_i+35(\frac{\sigma_i''}{2\sigma_i'}-\frac{2M_i'}{M_i})b_i+\frac{3\sigma_i''''}{8\sigma_i'}+\frac{35\sigma_i''M_i''}{4\sigma_i'M_i}+\frac{35M_i'''}{3M_i}\big)\chi_{i\hspace{0.15em}i\hspace{0.15em}i}^{111}\\
&
+\sum_{j\neq i}\frac{M_i}{\sigma_j'M_j}\big((3E_i+6A_i-\frac{\sigma_i''}{4\sigma_i'})(4E_j-a_j)H_{ij}-\frac{\sigma_i'(E_i+2A_i)+\frac{1}{4}\sigma_i''+\sigma_j'(4E_j-a_j)}{(q_i-q_j)^2}+\frac{2\sigma_i'}{(q_i-q_j)^3}\big)\chi_{i\hspace{0.15em}i\hspace{0.15em}i}^{111}\\
&
+\sum_{j\neq i}\big(12(E_j+3A_j-2a_j)\frac{H_{ij}^2}{\sigma_i'\sigma_j'}+(8E_i^2+(36A_i-37a_i)E_i+6A_i^2-15A_ia_i+9a_i^2+\frac{28}{3}b_i)\frac{M_jH_{ij}}{\sigma_i'M_i}\\
&\quad
-\frac{8H_{ij}}{\sigma_j'(q_i-q_j)^2}-\frac{(4E_i+2A_i-3a_i)\sigma_j'M_j}{\sigma_i'M_i(q_i-q_j)^2}+\frac{\sigma_j'M_j}{\sigma_i'M_i(q_i-q_j)^3}\big)\big[\chi_{i\hspace{0.05em}j\hspace{0.05em}j}^{111}+\mbox{perm}({}_3C_1)\big]\\
&
+\sum_{j,k\neq i}\frac{M_jH_{ij}}{\sigma_i'\sigma_k'M_k}\big((4E_k-a_k)H_{ik}-\frac{\sigma_i'}{3(q_i-q_k)^2}\big)\big[\chi_{i\hspace{0.05em}j\hspace{0.05em}j}^{111}+\mbox{perm}({}_3C_1)\big]+\sum_{j\neq i, k\neq j}\frac{4M_kH_{ij}^2H_{jk}}{\sigma_i'\sigma_j'^2M_j}\big[\chi_{i\hspace{0.05em}k\hspace{0.05em}k}^{111}+\mbox{perm}({}_3C_1)\big]\\
&
+\sum_{j\neq i, k\neq i,j}\frac{8H_{ij}H_{jk}H_{ki}}{\sigma_i'\sigma_j'\sigma_k'}\chi_{i\hspace{0.05em}j\hspace{0.05em}k}^{111}
+2\big(24E_i^2+(121A_i-75a_i)E_i-9A_i^2+25B_i+25b_i\big)\big[\chi_{i\hspace{0.15em}i\hspace{0.15em}i}^{112}+\mbox{perm}({}_3C_1)\big]\\
&
+\sum_{j \neq i}\frac{3M_i}{\sigma_j'M_j}\big((4E_j-a_j)H_{ij}-\frac{\sigma_i'}{3(q_i-q_j)^2}\big)\big[\chi_{i\hspace{0.15em}i\hspace{0.15em}i}^{112}+\mbox{perm}({}_3C_1)\big]+\sum_{j\neq i}\frac{12H_{ij}^2}{\sigma_i'\sigma_j'}\big[\chi_{i\hspace{0.05em}j\hspace{0.05em}j}^{112}+\mbox{perm}({}_3P_3)\big]\\
&
+\sum_{j\neq i}\frac{M_j}{\sigma_i'M_i}\big((12E_i+2A_i-3a_i)H_{ij}-\frac{\sigma_j'}{(q_i-q_j)^2}\big)\big[\chi_{i\hspace{0.05em}j\hspace{0.05em}j}^{211}+\mbox{perm}({}_3C_1)\big]\\
&
+5\big(\frac{13G}{\sigma_i'}+\frac{2\sigma_i''}{3\sigma_i'}+\frac{7M_i'}{M_i}\big)\big[\chi_{i\hspace{0.15em}i\hspace{0.15em}i}^{113}+\mbox{perm}({}_3C_1)\big]
+\sum_{j\neq i}\frac{5M_iH_{ij}}{\sigma_j'M_j}\big[\chi_{i\hspace{0.15em}i\hspace{0.05em}j}^{113}+\mbox{perm}({}_3C_1)\big]\\
&
+24\big(\frac{3G}{\sigma_i'}+\frac{M_i'}{M_i}\big)\big[\chi_{i\hspace{0.15em}i\hspace{0.15em}i}^{122}+\mbox{perm}({}_3C_1)\big]
+18\chi_{i\hspace{0.15em}i\hspace{0.15em}i}^{222}
+30\big[\chi_{i\hspace{0.15em}i\hspace{0.15em}i}^{123}+\mbox{perm}({}_3P_3)\big]+35\big[\chi_{i\hspace{0.15em}i\hspace{0.15em}i}^{114}+\mbox{perm}({}_3C_1)\big]\Big\}.\eqnum
\label{eqn:F8}
\end{align*}}

\begin{figure}[t]
 \begin{center}
  \includegraphics[width=100mm]{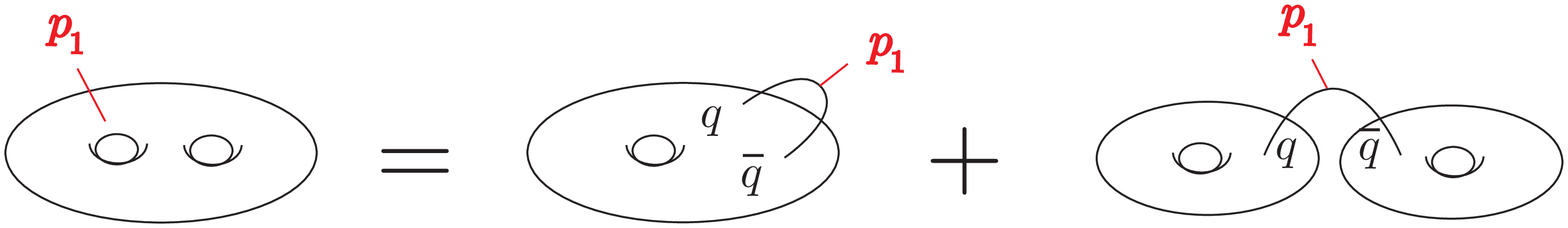}
 \end{center}
 \caption{Recursive structure of $W^{(2,1)}(p_1)$}
 \label{fig:F3}
\end{figure}
\noindent The meromorphic differential $W^{(2,1)}(p_1)$ is obtained as follows (see also Fig.\ref{fig:F3}):
{\footnotesize
\begin{align*}
&
W^{(2,1)}(p_1)=\sum_{q_i\in{\cal C}}\mathop{\mbox{Res}}_{q=q_i}\frac{dE_{q,\bar{q}}(p_1)}{2y(q)dq}\Big\{W^{(1,2)}(q,\bar{q})+W^{(1,1)}(q)W^{(1,1)}(\bar{q})\Big\} \\
&
=\frac{1}{64}\sum_{i}\frac{1}{\sigma_i'M_i^2}\Big\{\big(5E_i^4+(48A_i-55a_i)E_i^3+(\frac{381}{2}A_i^2-423A_ia_i+\frac{3769}{16}a_i^2-\frac{5}{2}B_i+\frac{191}{6}b_i)E_i^2\\
&\quad
+(135A_i^3-\frac{1767}{4}A_i^2a_i+(\frac{3975}{8}a_i^2-15B_i+211b_i)A_i-\frac{1521}{8}a_i^3+15a_iB_i-\frac{1399}{6}a_ib_i-\frac{31\sigma_i''''}{48\sigma_i'})E_i\\
&\quad
+\frac{417}{2}A_i^2b_i-466A_ia_ib_i-\frac{109\sigma_i''''}{96\sigma_i'}A_i+\frac{267\sigma_i''''}{192\sigma_i'}a_i+\frac{2109}{8}a_i^2b_i-5B_ib_i+\frac{190}{9}b_i^2\big)\chi_i^1\\
&
+\sum_{j\neq i}\big(\frac{M_i}{2\sigma_i'\sigma_j'^2M_j}(12E_i+24A_i-25a_i)H_{ij}^3+\frac{M_i^2}{2\sigma_j'^2M_j^2}(\frac{1}{8}(4E_j-a_j)^2+(8E_j+36A_j-49a_j)E_j+15A_j^2\\
&\quad
-30A_ja_j+21a_j^2-5B_j+\frac{31}{3}b_j)H_{ij}^2-\frac{25M_iH_{ij}^2}{6(q_i-q_j)^2\sigma_i'\sigma_j'M_j}-\frac{\sigma_i'M_i^2}{2(q_i-q_j)^2\sigma_j'^2M_j^2}(\frac{25}{3}E_j+4A_j-\frac{97}{12}a_j \\
&\quad +\frac{2}{q_i-q_j})H_{ij}+\frac{M_i}{2\sigma_j'M_j}(4E_j-a_j)(2E_i^2+(9A_i-\frac{37}{4}a_i)E_i+\frac{3}{2}A_i^2-\frac{15}{4}A_ia_i+\frac{9}{4}a_i^2+\frac{7}{3}b_i)H_{ij}\\
&\quad
+\frac{25\sigma_i'^2M_i^2}{144(q_i-q_j)^4\sigma_j'^2M_j^2}-\frac{\sigma_i'M_i}{6(q_i-q_j)^2\sigma_j'M_j}(E_i^2+6(A_i-a_i)E_i+2b_i)+\frac{\sigma_i'M_i}{8(q_i-q_j)^2\sigma_j'M_j}(-\frac{4}{3}E_i^2 \\
&\quad
+(12A_i-\frac{35}{3}a_i)E_i-3A_i^2+3A_ia_i+\frac{14}{9}b_i-(4E_i+2A_i-3a_i)(4E_j-a_j)\frac{\sigma_j'}{\sigma_i'}+\frac{1}{q_i-q_j}(8E_i-12A_i \\
&\quad
+10a_i+(4E_j-a_j)\frac{\sigma_j'}{\sigma_i'})-\frac{5}{(q_i-q_j)^2})\big)\chi_i^1 \\
&
+\sum_{j\neq i, k\neq i,j}\frac{M_i^2}{\sigma_j'^2\sigma_k'^2M_jM_k}\big((4H_{ik}H_{jk}^2+\sigma_k'(4E_k-a_k)H_{ij}H_{jk}-\frac{\sigma_j'\sigma_k'H_{ij}}{3(q_j-q_k)^2})\frac{H_{ij}}{2} \\
&\quad
+\sigma_j'\sigma_k'((E_j-\frac14a_j)H_{ij}-\frac{\sigma_i'}{12(q_i-q_j)^2})((E_k-\frac14a_k)H_{ik}-\frac{\sigma_i'}{12(q_i-q_k)^2})\big)\chi_i^1 \\
&
+\big(10E_i^3+(61A_i-75a_i)E_i^2+(45A_i^2-\frac{409}{4}A_ia_i+\frac{507}{8}a_i^2-5B_i+31b_i)E_i+\frac{121}{3}A_ib_i-\frac{97}{2}a_ib_i-\frac{73\sigma_i''''}{384\sigma_i'}\big)\chi_i^2 \\
&
+\sum_{j\neq i}\frac{M_i}{\sigma_j'M_j}\big(\frac{4H_{ij}^3}{\sigma_i'\sigma_j'}+(\frac{3}{2}E_i+\frac14A_i-\frac38a_i)(4E_j-a_j)H_{ij}+\frac{\sigma_i'}{4(q_i-q_j)^3} \\
&\quad
-\frac{\sigma_i'}{(q_i-q_j)^2}(\frac12E_i-\frac{5}{12}A_i+\frac38a_i+\frac{(4E_j-a_j)\sigma_j'}{8\sigma_i'})\big)\chi_i^2 \\
&
+\big(\frac{35}{2}E_i^2+(\frac{35}{2}A_i-\frac{245}{8}a_i)E_i+\frac{49}{16}A_i^2-\frac52B_i+\frac{245}{24}b_i\big)\chi_i^3
+\sum_{j\neq i}\frac{M_i}{\sigma_j'M_j}\big(\frac58(4E_j-a_j)H_{ij}-\frac{5\sigma_i'}{24(q_i-q_j)^2}\big)\chi_i^3 \\
&
+\big(\frac{35}{2}E_i-\frac{49}{8}A_i\big)\chi_i^4+\frac{105}{16}\chi_i^5\Big\}.\eqnum
\label{eqn:F9}
\end{align*}}

\subsection{Figure eight knot complement}
\baselineskip 17.5pt

The free energies (\ref{eqn:3.45}) with $\chi=-3$ on the curve (\ref{eqn:3.57}), (\ref{eqn:3.58}) for the figure eight knot complement are summarized as follows:\vspace{-1em}

{\footnotesize
\begin{align*}
\label{eqn:F10}
&
{\cal F}^{(0,5)}(p)=\frac{1}{{\widetilde \sigma}(w)^{9/2}}\Big(-\frac{4}{3}w^8+\frac{10}{3}w^7-3w^6-\frac{119}{6}w^5+\frac{2273}{60}w^4-\frac{997}{120}w^3-\frac{9271}{240}w^2+\frac{6357}{160}w-\frac{1411}{120}\Big), \eqnum\\
&
{\cal F}^{(1,3)}(p)=-\frac{(4w-3)^3G_3}{162000{\widetilde \sigma}(w)^{3/2}}-\frac{(4w-3)G_2}{300{\widetilde \sigma}(w)^{5/2}}\Big(\frac{16}{27}w^4-\frac{58}{27}w^3+\frac{13}{9}w^2+\frac{7}{2}w-\frac{9}{4}\Big) \\
&
-\frac{G_1}{{\widetilde \sigma}(w)^{7/2}}\Big(\frac{2368}{30375}w^7-\frac{17714}{30375}w^6+\frac{29443}{30375}w^5+\frac{106}{81}w^4-\frac{1267}{450}w^3-\frac{3239}{9000}w^2+\frac{21823}{6000}w-\frac{3637}{2000}\Big) \\
&
-\frac{1}{{\widetilde \sigma}(w)^{9/2}}\Big(\frac{15616}{54675}w^9-\frac{60448}{18225}w^8+\frac{139624}{18225}w^7+\frac{103012}{10935}w^6-\frac{19846}{2025}w^5-\frac{23747}{675}w^4+\frac{85901}{1350}w^3\\
\label{eqn:F11}
&\hspace{5em}
+\frac{4151}{300}w^2-\frac{3277}{40}w+\frac{140863}{3600}\Big), \eqnum\\
&
{\cal F}^{(2,1)}(p)=-\frac{4(4w-3){\widetilde G}_3}{50625{\widetilde \sigma}(w)^{1/2}}
+\frac{{\widetilde G}_2}{900{\widetilde \sigma}(w)^{5/2}}\Big(\frac{2944}{135}w^5-\frac{8807}{135}w^4+\frac{1003}{54}w^3+\frac{13813}{180}w^2-\frac{559}{24}w-\frac{85}{2}\Big) \\
&
+\frac{{\widetilde G}_1}{1125{\widetilde \sigma}(w)^{7/2}}\Big(\frac{88448}{405}w^7-\frac{24656}{27}w^6+\frac{106688}{135}w^5+\frac{349636}{405}w^4-\frac{104866}{135}w^3-\frac{6142}{5}w^2-\frac{187}{30}w+\frac{11311}{10}\Big) \\
&
+\frac{1}{5625{\widetilde \sigma}(w)^{9/2}}\Big(\frac{827392}{243}w^9-\frac{4380472}{243}w^8+\frac{1967444}{81}w^7+\frac{3066698}{243}w^6-\frac{11109413}{243}w^5+\frac{13223}{54}w^4 \\
\label{eqn:F12}
&\hspace{5em}
-\frac{2184931}{108}w^3+\frac{3318629}{72}w^2+\frac{786203}{16}w-\frac{451859}{8}\Big),\eqnum
\end{align*}}
\vspace{-1em}

\noindent where by distinguishing $G$ in ${\cal F}^{(1,3)}(p)$ from $G$ in ${\cal F}^{(2,1)}(p)$, we put $G_1=G,~G_2=G^2,~G_3=G^3$ in \ref{eqn:F11}, and ${\widetilde G}_1=G,~{\widetilde G}_2=G^2,~{\widetilde G}_3=G^3$ in \ref{eqn:F12}. If we regularize these parameters as
\begin{equation}
G_1=\frac{7}{3},\quad G_2=-\frac{95}{9},\quad G_3=\Big(\frac{7}{3}\Big)^3,\quad {\widetilde G}_1=\frac{7}{3},\quad {\widetilde G}_2=-\frac{47}{9},\quad {\widetilde G}_3=-\frac{131}{27},
\label{eqn:F13}
\end{equation}
then the free energy
\begin{eqnarray}
{\cal F}_4(p)&=&4\big({\cal F}^{(0,5)}(p)+{\cal F}^{(1,3)}(p)+{\cal F}^{(2,1)}(p)\big)\nonumber\\
&=&
\frac{1}{{15\widetilde \sigma}(w)^{9/2}}\Big(\frac{128}{3}w^8-\frac{256}{3}w^7-\frac{4352}{3}w^6+\frac{1024}{3}w^5+\frac{14896}{3}w^4\nonumber\\
&&\hspace{5em}
-\frac{23464}{3}w^3+212w^2+8194w-\frac{27469}{6}\Big)
\label{eqn:F14}
\end{eqnarray}
coincides with the perturbative invariant (\ref{S4Fig}), where $w=(m^2+m^{-2})/2$.

\subsection{Once punctured torus bundle over $\mathbb{S}^1$ with holonomy $L^2 R$}
\baselineskip 17.5pt

The free energies (\ref{eqn:3.45}) with $\chi=-3$ on the curve
(\ref{eqn:3.71}), (\ref{eqn:3.72}) for the once punctured torus bundle
over $\mathbb{S}^1$ with holonomy $L^2 R$ are summarized as
follows:\vspace{-1em}

{\footnotesize
\begin{align*}
\label{eqn:F15}
&
{\cal F}^{(0,5)}(p)=\frac{1}{{\widetilde \sigma}(w)^{9/2}}\Big(-\frac{4}{3}w^8-\frac{58}{3}w^6-\frac{23}{2}w^5+\frac{319}{20}w^4-\frac{2069}{60}w^3-\frac{47}{12}w^2+\frac{13651}{480}w-\frac{5341}{480}\Big), \eqnum\\
&
{\cal F}^{(1,3)}(p)=-\frac{(6w-7)^3G_3}{1053696{\widetilde \sigma}(w)^{3/2}}-\frac{(6w-7)G_2}{784{\widetilde \sigma}(w)^{5/2}}\Big(\frac{3}{7}w^4-\frac{17}{7}w^3+\frac{17}{6}w^2+\frac{71}{12}w-\frac{133}{48}\Big) \\
&
-\frac{G_1}{{8\widetilde \sigma}(w)^{7/2}}\Big(\frac{45}{343}w^7-\frac{1727}{686}w^6+\frac{27595}{4116}w^5+\frac{42659}{3528}w^4-\frac{181607}{7056}w^3+\frac{33499}{2016}w^2+\frac{134411}{4032}w-\frac{20671}{1152}\Big) \\
&
+\frac{1}{{\widetilde \sigma}(w)^{9/2}}\Big(\frac{89}{1029}w^9-\frac{1515}{686}w^8+\frac{8870}{1029}w^7+\frac{956534}{27783}w^6-\frac{77521}{10584}w^5+\frac{38095}{784}w^4+\frac{37477}{324}w^3\\
\label{eqn:F16}
&\hspace{5em}
-\frac{35419}{336}w^2-\frac{558557}{6912}w+\frac{2672231}{41472}\Big), \eqnum\\
&
{\cal F}^{(2,1)}(p)=\frac{81(6w-7){\widetilde G}_3}{2458624{\widetilde \sigma}(w)^{1/2}}
+\frac{{\widetilde G}_2}{224{\widetilde \sigma}(w)^{5/2}}\Big(\frac{981}{343}w^5-\frac{6101}{686}w^4-\frac{3911}{686}w^3+\frac{6605}{196}w^2-\frac{471}{112}w-\frac{1063}{32}\Big) \\
&
-\frac{{\widetilde G}_1}{56{\widetilde \sigma}(w)^{7/2}}\Big(\frac{1629}{343}w^7-\frac{16263}{686}w^6+\frac{20913}{1372}w^5+\frac{194077}{2744}w^4-\frac{53455}{784}w^3-\frac{42725}{224}w^2+\frac{4253}{64}w+\frac{21273}{128}\Big) \\
&
+\frac{1}{98{\widetilde \sigma}(w)^{9/2}}\Big(\frac{1115}{49}w^9-\frac{123383}{2205}w^8+\frac{123383}{2205}w^7+\frac{2396633}{4410}w^6-\frac{24202151}{17640}w^5-\frac{2692853}{720}w^4 \\
\label{eqn:F17}
&\hspace{5em}
+\frac{751187}{720}w^3+\frac{8009701}{1440}w^2-\frac{6519401}{11520}w-\frac{14175847}{4608}\Big),\eqnum
\end{align*}}
\vspace{-1em}

\noindent where as in the case of the figure eight knot complement, we
distinguished $G$ in ${\cal F}^{(1,3)}(p)$ from $G$ in ${\cal
  F}^{(2,1)}(p)$. In this example, as in (\ref{S2_m009}) the partition
function for the state integral model contains imaginary
terms. Therefore we have to add the imaginary term to our result for
the comparison with (\ref{S4_m009}), but we do not find natural
regularizations for the above $G$'s, and natural choice of the
imaginary term. We leave the problem to future work.

\end{document}